\documentclass[aps,prb,amsmath,amssymb,reprint]{revtex4-1}

\usepackage{graphicx}
\usepackage{url}
\usepackage{amsthm}
\usepackage[version=3]{mhchem}
\usepackage[utf8]{inputenc}
\usepackage[tight]{units}
\usepackage{bm}

\usepackage{algorithm}
\usepackage{algpseudocode}

\newcommand{\norm}[1]{\left\|#1\right\|}

\DeclareMathOperator*{\argmin}{argmin}
\newtheorem{lemma}{Lemma}

\newtheorem{theorem}{Theorem}
\usepackage{todonotes}

\def\R{{\mathbb R}}
\def\Tr{{\rm Tr}}
\DeclareMathOperator{\diag}{diag}
\DeclareMathOperator{\diagm}{diagm}
\newcommand{\abs}[1]{\left|#1\right|}

\usepackage{todonotes}

\newcommand{\E}{{\mathcal E}}

\newcommand{\I}{{\mathcal I}}
\newcommand{\HKS}{H_{\rm KS}}

\newcommand{\ffd}{f_{\rm FD}}
\newcommand{\dVn}{\delta V_n}

\begin{document}

\title{
    A robust and efficient line search for self-consistent field iterations
}
\author{Michael F. Herbst}
\email{herbst@acom.rwth-aachen.de}
\affiliation{
    Applied and Computational Mathematics, RWTH Aachen University, Schinkelstr. 2, 52062 Aachen, Germany.
}

\author{Antoine Levitt}
\email{antoine.levitt@inria.fr}
\affiliation{
  Inria Paris and CERMICS, \'Ecole des Ponts, 6 \& 8 avenue Blaise Pascal, 77455 Marne-la-Vall\'ee, France.
}

\begin{abstract}
  We propose a novel adaptive damping algorithm for the
  self-consistent field (SCF) iterations of Kohn-Sham
  density-functional theory, using a backtracking line search to
  automatically adjust the damping in each SCF step. This line search
  is based on a theoretically sound, accurate and inexpensive model
  for the energy as a function of the damping parameter. In contrast
  to usual SCF schemes, the resulting algorithm is fully automatic
  and does not require the user to select a damping. We
  successfully apply it to a wide range of challenging systems,
  including elongated supercells, surfaces and transition-metal alloys.
\end{abstract}
\maketitle

\section{Introduction}
\textit{Ab initio} simulation methods
are standard practice for predicting the
chemical and physical properties of molecules and materials.
At the level of simulating electronic structure the
majority of approaches are either directly based
upon Kohn-Sham density-functional theory~(DFT) or Hartree-Fock~(HF)
or use these techniques as starting points for more accurate
post-DFT or post-HF developments.
Both HF and DFT ground states are commonly found by solving
the self-consistent field~(SCF) equations,
which for both type of methods are very similar in structure.
Being thus fundamental to electronic-structure simulations
substantial effort has been devoted in the past to develop
efficient and widely applicable SCF algorithms.
We refer to \citet{Woods2019} and \citet{Lehtola2020}
for recent reviews on this subject.

However, the advent of both cheap computational power
as well as the introduction of data-driven approaches to materials modelling
has caused simulation practice to change noticeably.
In particular in domains such as catalysis or battery research
where experiments are expensive or time-consuming,
it is now standard practice
to perform systematic computations on thousands to millions of compounds.
The aim of such high-throughput calculations is to either
(i) generate data for training sophisticated surrogate models
or to (ii) directly screen complete design spaces for relevant compounds.
The development of such data-driven strategies has already accelerated
research in these fields and enabled
the discovery of novel semiconductors, electrocatalysts,
materials for hydrogen storage or for Li-ion batteries%
~\cite{Jain2016,Alberi2019,Luo2021}.

Compared to the early years where the aim was to perform
a small number of computations on hand-picked systems,
high-throughput screening approaches have much stronger requirements.
In particular the key bottleneck is the required human time
to set up and supervise computations.
To minimize manual effort state-of-the-art high-throughput frameworks%
~\cite{Curtarolo2012, %
Jain2011, %
Huber2020} %
provide a set of heuristics
which automatically select computational parameters based on prior experience.
In case of a failing calculation such heuristics may also be employed
for parameter adjustment and automatic rescheduling.
While this empirical approach manages to take care of the majority of failures
automatically, it is far from perfect.
First, state-of-the-art heuristic approaches cannot capture all cases,
and keeping in mind the large absolute number of calculations already a 1\% fraction
of cases that require human attention easily equals hundreds to thousands of calculations.
This causes idle time and severely limits the overall throughput of a study.
Second, any failing calculation, whether automatically caught
by a high-throughput framework or not, needs to be redone, implying
wasted computational resources that contributes to the
already noteworthy environmental footprint of supercomputing~\cite{Feng2007,Feng2008}.
The objectives for improving the algorithms employed in high-throughput workflows
is therefore to increase the inherent reliability as well as reduce the number of
parameters, which need to be chosen.
Ideally each building block of a simulation workflow would be entirely black-box
and automatically self-adapt to each simulated system.
To some extent this amounts to taking the existing empirical wisdom
already implemented in existing high-throughput frameworks
and converting it into simulation algorithms with convergence guarantees
using a mixture of both mathematical and physical arguments.

With this objective in mind, this work will focus on improving the robustness
of self-consistent field~(SCF) algorithms,
as mentioned above
one of the most fundamental components of electronic-structure simulations.
Our main motivation
and application are DFT simulations discretized in plane wave or
``large'' basis sets, for which it is only feasible to store and
compute with orbitals, densities and potentials, and not the full
density matrix or Fock matrix.
In this setting, the standard SCF approach are damped, preconditioned self-consistent iterations.
Using an approach based on potential-mixing
the next SCF iterate is found as
\begin{equation}
    V_\text{next} = V_\text{in} + \alpha P^{-1} (V_\text{out} - V_\text{in}),
    \label{eqn:potmixsimple}
\end{equation}
where $V_\text{in}$ and $V_\text{out}$ are the input and output potentials
to a simple SCF step, $\alpha$ is a fixed damping parameter and $P$ is a preconditioner.
It is well-known that simple SCF iterations (where $P$ is the identity)
can converge poorly for many systems due to a number of instabilities~\cite{ldos}.
Examples are the large-wavelength divergence due to the Coulomb operator
leading to the ``charge-sloshing'' behavior in metals
or the effect of strongly localized states near the Fermi level,
e.g.~due to surface states or $d$- or $f$-orbitals.
To accelerate the convergence of the SCF iteration despite these instabilities,
one typically aims to employ a preconditioner $P$ matching the underlying
system.
Despite some recent progress towards cheap self-adapting
preconditioning strategies~\cite{ldos}
for the charge-sloshing-type instabilities,
choosing a matching preconditioner is still not a straightforward task
for other types of instabilities.
For example currently no cheap preconditioner is available to
treat the instabilities due to strongly localized states near the Fermi level,
such that in such systems using a suboptimal preconditioning strategy is unavoidable.
While convergence acceleration techniques are usually crucial in such cases,
these also complicate the choice of an appropriate damping parameter $\alpha$
to achieve the fastest and most reliable convergence.
As we will detail in a series of example calculations
on some transition metal systems the interplay of mismatching preconditioner
and convergence acceleration can lead to a very unsystematic pattern
between the chosen damping parameter $\alpha$ and obtaining a successful
or failing calculation.
Especially for such cases finding a good combination of preconditioning strategy
and damping parameter can require substantial trial and error.

As an alternative approach to a fixed damping selected by a user
\textit{a priori} Cancès and Le Bris suggested the optimal damping
algorithm~(ODA)~\cite{Cances2000,Cances2000a}. In this algorithm the
damping parameter is obtained automatically by performing a line
search along the update suggested by a simple SCF step. Following this
strategy, the ODA ensures a monotonic decrease of the energy, which
leads to strong convergence guarantees. This can be improved using the
history to improve convergence, such as in the EDIIS
method~\cite{Kudin2002}, or trust-region strategies
\cite{francisco2004globally,francisco2006density}. These approaches
are successfully employed for SCF calculations on atom-centered basis
sets, where an explicit representation of the density matrix is
possible. However, their use with plane-wave DFT methods, where only
orbitals, densities and potentials are ever stored, does not appear to
be straightforward, in particular in conjunction with accelerated
methods.

Another development towards finding an DFT ground state in a
mathematically guaranteed fashion are approaches based on a direct
minimization of the DFT energy as a function of the orbitals and
occupations, not using the self-consistency principle (see Reference
\onlinecite{cances2020convergence} for a mathematical comparison).
Although direct minimization methods are often quite efficient for
gapped systems, their use for metals requires a minimization over
occupation numbers \cite{marzari1997ensemble,freysoldt2009direct},
which is potentially costly and unstable. For this reason such
approaches seem to be less used than the SCF schemes in solid-state
physics.

In the realm of self-consistent iterations, variable-step methods have
been successfully used \cite{Marks2021,marks2008robust} to increase
robustness. These methods are based on a minimization of the residual.
Although this often proves efficient in practice, this has a number of
disadvantages. First, the residual might go up then down on the way to
a solution making it rather hard to design a linesearch algorithm.
Second, this forces an algorithm to select an appropriate notion of a
residual norm, with results potentially sensitive to this choice.
Third, there is the possibility of getting stuck in local minima of
the residual, or a saddle point of the energy. By contrast, we aim to
find a scheme ensuring energy decrease as an important ingredient to
ensure robustness. Indeed, under mild conditions, a scheme that
decreases the energy monotonically is guaranteed to converge to a
solution of the Kohn-Sham equations (see Theorems 1 and 2 below). This
is in contrast to residual-based schemes, which afford no such
guarantee. The very good practical performance of these schemes,
despite the lack of global theoretical guarantees, is an interesting
direction for future research.

Our goal in this work is to design a mixing scheme that (a) is
applicable to plane-wave DFT, and involves only quantities such as
densities and potentials; (b) is based on an energy minimization, to
ensure robustness; (c) is based on the self-consistent iterations; (d)
is compatible with acceleration and preconditioning. Our scheme is based on a minimal modification of the damped
preconditioned iterations \eqref{eqn:potmixsimple}. Similar to the ODA
approach we employ a line search procedure to choose the damping
parameter automatically. Our algorithm builds upon ideas of the
potential-based algorithm of \citet{gonze1996towards} to construct an
efficient SCF algorithm. In combination with Anderson acceleration on
challenging systems we show our adaptive damping scheme to be less
sensitive than the approach based on a fixed damping parameter.
In contrast to the fixed damping approach the scheme does
not require a manual damping selection from the user.

The outline of the paper is as follows.
Section \ref{sec:analysis} presents
the mathematical analysis of the self-consistent field iterations
justifying our algorithmic developments.
In particular it presents a justification for global convergence
of the SCF iterations. The proofs for the results presented in this section are given in the appendix.
Section \ref{sec:adaptive} discusses the adaptive damping algorithm itself
followed by numerical tests (Section \ref{sec:tests}) to illustrate
and contrast cost and performance compared to the standard fixed-damping approach.
Concluding remarks and some outlook to future work is given in Section \ref{sec:conclusion}.

\section{Analysis}
\label{sec:analysis}
\subsection{Preliminaries}
\label{sec:prelim}
We use similar notation to those in \citet{cances2020convergence},
extend the analysis in that paper to the finite-temperature
case~\cite{mermin1965thermal}, and
introduce the potential mixing algorithm. We
work in the \textit{grand-canonical ensemble}: we fix a chemical
potential (or Fermi level) $\mu$ and an inverse temperature $\beta$.
In particular, the number of electrons is not fixed. This is for
mathematical convenience: fixing the number of electrons $N$ instead
of $\mu$ does not change our results. We assume that space has been
discretized in a finite-dimensional orthogonal basis (typically,
plane-waves) of size $N_{\rm b}$, and will not treat either spin or
Brillouin zone sampling explicitly for notational simplicity, although
of course the formalism can be extended easily. In this section we
will work with the formalism of density matrices, self-adjoint
operators $P$ satisfying $0 \le P \le 1$. Such operators can be
diagonalized as
\begin{equation}
  P = \sum_{i=1}^{N_{\rm b}} f_{i} |\phi_{i}\rangle\langle \phi_{i}|.
\end{equation}
The numbers $0 \le f_{i} \le 1$ are the occupation numbers, and $\phi_{i}$
are the orbitals. Either density matrices or the set of occupation
numbers and orbitals can be taken as the primary unknowns in the
self-consistency problem. Density matrices are impractical numerically
in plane-wave basis sets,
since they are $N_{\rm b} \times N_{\rm b}$; however,
they are very convenient to formulate and analyze algorithms.
Accordingly, we will use them in this theoretical section, but
implement the resulting algorithms using orbitals only.

We work on the sets
\begin{align}
  \mathcal H &= \{H \in \R^{N_{\rm b} \times N_{\rm b}}, H^{T} = H\}\\
  \mathcal P &= \{P \in \mathcal H, 0 < P < 1\}
\end{align}
of Hamiltonians and density matrices, equipped with the standard
Frobenius metric. Here and in the following, inequalities between matrices are
understood in the sense of symmetric matrices. The closure
\mbox{$\overline{\mathcal P} = \{P \in \mathcal H, 0 \le P \le 1\}$} is
compact. Let $\E_{0}$ be a twice continuously differentiable function on
$\overline{\mathcal P}$: we aim to solve the problem
\begin{align}
  \min_{P \in \mathcal P} \E_0(P).
\end{align}
Let
\begin{align}
  \HKS(P) = \nabla\E_{0}(P)
\end{align}
be its gradient, and
\begin{align}
  \label{eqn:kernel4}
  \bm K(P) = \bm{d^{2}\E_{0}}(P) = \bm{d \nabla\E_{0}}(P)
\end{align}
be its Hessian. We will denote in bold ``super-operators'' or ``four-point operators'', operators from $\mathcal H$ to
$\mathcal H$. Let $s$ be the fermionic entropy
\begin{align}
  s(p) &= -(p\log p + (1-p)\log (1-p)),
\end{align}
with derivatives
\begin{align}
    s'(p) &= \log\left( \frac {1-p}{p} \right), \quad s''(p) = -\frac{1}{p(1-p)}.
\end{align}
Let
\begin{align}
  \E(P) &= \E_{0}(P) - \frac 1 \beta \Tr(s(P)) - \mu \Tr P.
\end{align}
be the free energy of a density matrix, where here and in the
following we use functional calculus implicitly to define $s(P) \in
\mathcal H$.
$\E$ diverges on the boundary of $\mathcal P$, whose closure is
compact, and therefore $\E$ has at least one minimizer in
$\mathcal P$. The first-order optimality condition
$\nabla \E(P) = 0$ gives
\begin{align}
  \HKS(P) -\mu - \frac 1 \beta s'(P) = 0,
\end{align}
and therefore
\begin{align}
    \label{eqn:scfdm}
    P = \ffd(\HKS(P)),
\end{align}
where we define the Fermi-Dirac map $\ffd$ by
\begin{align}
    \label{eqn:fermidirac}
  \ffd(H) = \frac{1}{1+e^{\beta(H-\mu)}}.
\end{align}
Here we have used the equation
$s'(\ffd(\varepsilon)) = \beta(\varepsilon - \mu)$, which will also be
useful in the following. Although we use the Fermi-Dirac smearing
function for concreteness, our results apply just as well to Gaussian
smearing for instance; however, they don't apply to schemes with
non-monotonous occupations such as the Methfessel-Paxton
scheme \cite{methfessel1989high}.

\subsection{The dual energy}
Reformulating the ideas in \citet{gonze1996towards}, we now define a ``dual'' energy
\begin{align}
  \I(H) &= \E(\ffd(H)).
\end{align}
Since the map $\ffd$ is a
bijection from $\mathcal H$ to $\mathcal P$, we have
\begin{align}
  \min_{H \in \mathcal H} \I(H) = \min_{P \in \mathcal P} \E(P).
\end{align}
This is analogous to convex duality since the unknown in this
formulation is now $H = \nabla \E(P)$.

We can compute the derivative $\bm{\chi_{0}}= d \ffd$ of $\ffd$ (see
Lemma \ref{sec:lemma} in the Appendix for details):
\begin{equation}
    \label{eqn:chi4}
\begin{aligned}
  &\bm{\chi_{0}}\left(\sum_{i=1}^{N_{\rm b}} \varepsilon_{i} |\phi_{i}\rangle\langle \phi_{i}|\right) \cdot \delta H \\
  &\hspace{1.9em}= \sum_{i=1}^{N_{\rm b}}\sum_{j=1}^{N_{\rm b}} \frac{\ffd(\varepsilon_{i}) - \ffd(\varepsilon_{j})}{\varepsilon_{i}-\varepsilon_{j}}\langle  \phi_{i}, \delta H \phi_{j} \rangle |\phi_{i}\rangle\langle \phi_{j}|
\end{aligned}
\end{equation}
The linear map $\bm{\chi_{0}}$ is a ``four-point'' generalization of the
independent-particle polarizability. It describes the change to the
density matrix of a system of independent electrons to a change in Fock matrix.

We then have
\begin{equation}
    \begin{aligned}
      \nabla \I(H) &= \bm{\chi_{0}}(H) \nabla \E(\ffd(H)) \\
      &= \bm{\chi_{0}}(H) \Big(  \HKS(\ffd(H)) -\mu - \frac 1 \beta s'(\ffd(H))\Big)\\
      &= \bm{\chi_{0}}(H) (  \HKS(\ffd(H)) -H)
    \end{aligned}
    \label{eqn:gradI}
\end{equation}
where again we used $s'(\ffd(\varepsilon)) = \beta(\varepsilon - \mu)$.

The Hessian of $\I$ is a complicated object due to the derivative of
$\bm{\chi_{0}}(H)$. However, at a solution of $\HKS(\ffd(H_{*})) = H_{*}$,
this term vanishes, and we have the simple result
\begin{equation}
    \label{eqn:hessI}
    \bm{d^{2} \I}(H_{*}) = -\bm{\chi_{0}}(H_{*}) (1- \bm K(H_{*}) \bm{\chi_{0}}(H_{*}))
\end{equation}

To better understand this object, we compute the Hessian of $\E$.
From $\nabla \E(P) = \HKS(P) - \frac 1 \beta s'(P) - \mu$ we get
\begin{equation}
\begin{aligned}
  \bm{d^{2} \E}(P) \cdot \delta P
  &= \bm K(\HKS(P)) \cdot \delta P \\&- \frac 1 \beta \sum_{i=1}^{N_{\rm b}}\sum_{j=1}^{N_{\rm b}} \frac{s'(p_{i})-s'(p_{j})}{p_{i}-p_{j}}\langle  \phi_{i}, \delta P \phi_{j} \rangle |\phi_{i}\rangle\langle \phi_{j}|.
\end{aligned}
\end{equation}
Defining
\begin{equation}
\begin{aligned}
  &\bm \Omega\left(\sum_{i=1}^{N_{\rm b}} \varepsilon_{i} |\phi_{i}\rangle\langle \phi_{i}|\right) \cdot \delta P\\
  &\hspace{2.5em}= -\sum_{i=1}^{N_{\rm b}}\sum_{j=1}^{N_{\rm b}} \frac{\varepsilon_{i}-\varepsilon_{j}}{\ffd(\varepsilon_{i}) - \ffd(\varepsilon_{j})}\langle  \phi_{i}, \delta P \phi_{j} \rangle |\phi_{i}\rangle\langle \phi_{j}|
\end{aligned}
\end{equation}
we get
\begin{align}
  \label{eqn:hessE}
  \bm{d^{2} \E}(P) &= \bm K(\HKS(P)) + \bm \Omega(\ffd^{-1}(P)).
\end{align}
The point of this formula is to recognize now that $\bm \Omega(H) = -\bm{\chi_{0}}(H)^{-1}$. This links the Hessians of $\E$ and $\I$: at
a fixed point $H_{*} = \HKS(\ffd(H_{*}))$,
\begin{align}
  \label{eqn:hess_both}
  \bm{d^{2}\E}(\ffd(H_{*})) = \bm \Omega(H_{*}) {\bm{d^{2}\I}}(H_{*}) \bm \Omega(H_{*}).
\end{align}
Since $\bm \Omega$ is self-adjoint and positive definite, both Hessians
have the same inertia (number of negative eigenvalues). 

\subsection{Hamiltonian mixing}
\label{sec:hammix}
The very simplest Hamiltonian mixing algorithm is
\begin{align}
  H_{n+1} = \HKS(\ffd(H_{n})).
\end{align}
As already recognized in Reference \onlinecite{gonze1996towards}, \eqref{eqn:gradI}
makes it possible to reinterpret this simple algorithm in a new light:
it is a gradient descent algorithm on $\I$ with step $1$,
preconditioned by $\bm{\chi_{0}}(H_{n})^{-1}$. It is natural to use a
smaller stepsize to try to ensure convergence, and indeed this is
guaranteed to work:
\begin{theorem}
  \label{thm:algorithm}
  Let $H_{0} \in \mathcal H$. There is $\alpha_{0} > 0$ such that, for
  all $0 < \alpha < \alpha_{0}$, the algorithm
  \begin{align}
    H_{n+1} = H_{n} + \alpha(\HKS(\ffd(H_{n})) - H_{n})
  \end{align}
  satisfies $\HKS(\ffd(H_{n})) - H_{n} \to 0$. If furthermore $E$ is
  analytic, $H_{n}$ converges to a solution of the equation $\HKS(\ffd(H))=H$.
\end{theorem}
Adaptive-step schemes can also ensure guaranteed convergence:
\begin{theorem}
  \label{thm:damping}
  Fix $H_{0} \in \mathcal H$, and constants $0 < \alpha_{\rm max} < 1$,
  $0 < c < 1, 0 < \tau < 1$. Consider the algorithm
  \begin{align}
    H_{n+1} = H_{n} + \alpha_{n}(\HKS(\ffd(H_{n})) - H_{n})
  \end{align}
  where $\alpha_{n}$ is chosen in the following way: starting from
  $\alpha_{\rm max}$, decrease $\alpha_{\rm max}$ by a factor $\tau$
  while the Armijo line search condition
  \begin{equation}
    \label{eq:armijo}
      \begin{aligned}
    &\I(H_{n} + \alpha_{n}(\HKS(\ffd(H_{n})) - H_{n})) \\
    &\hspace{1.0em}\le \I(H_{n}) - \alpha_{n} c \langle  \bm \Omega(\ffd(H_{n})) \nabla\I(H_{n}), \nabla\I(H_{n}) \rangle
      \end{aligned}
  \end{equation}
  is not verified. Then this algorithm satisfies $\HKS(\ffd(H_{n})) - H_{n} \to 0$. If furthermore $E$ is
  analytic, $H_{n}$ converges to a solution of the equation $\HKS(\ffd(H))=H$.
\end{theorem}
The proofs of both these statements are found in the Appendix.

The adaptive-step scheme above however suffers from two important
drawbacks. First, it is costly (requiring several SCF steps per
iteration). Second, it is imcompatible with preconditioned or
accelerated schemes because, in contrast to the SCF direction, there
is no guarantee in these cases that the chosen direction is a descent
direction to the energy. This would make a straightforward
implementation of the above algorithm uncompetitive for ``easy''
systems, and therefore motivates the search for a compromise algorithm
that tries to recover some robustness properties while not sacrificing
performance.

\subsection{Potential mixing}
We now specialize the above discussion to our case of interest of semi-local
density-functional theory~(DFT) models. We introduce the operators
\mbox{$\diag : \R^{N_{\rm b} \times N_{\rm b}} \to \R^{N_{\rm b}}$} and
\mbox{$\diagm: \R^{N_{\rm b}} \to \R^{N_{\rm b} \times N_{\rm b}}$}. The
$\diag$ operator takes the diagonal (in real space) of a density
matrix, yielding a density. The $\diagm$ operator constructs a Fock
matrix contribution with a given local potential.
Both operators are adjoint of each other.
With these notations, the energy function takes the form
\begin{equation}
    \E_0(P) = \Tr(H_0 P) + g\left(\diag(P)\right),
    \label{eqn:trace}
\end{equation}
where $H_0$ is a given operator (the core Hamiltonian) and $g$ is a
nonlinear function (the Hartree-exchange-correlation energy). For these models the gradient
of $\E_0(P)$ (the Fock matrix) depends on $\diag(P)$ (the density) only:
\begin{equation}
    H_\text{KS}(P) = H_0 + \diagm(V(\diag(P))),
\end{equation}
with the potential
\begin{equation}
    V(\rho) = \nabla g(\rho) \in \R^{N_{\rm b}}.
\end{equation}
Based on \eqref{eqn:fermidirac} and the definition of the density
we define the potential-to-density mapping
\begin{equation}
  \rho(V) =  \diag(\ffd(H_0 + \diagm(V))),
    \label{eqn:potdensmap}
\end{equation}
which allows to solve the self-consistency problem
\eqref{eqn:scfdm} by iteration in the potential $V$ only:
\newcommand{\Vsearch}{\delta V_n}
\begin{equation}
    \label{eqn:potmix}
    V_{n+1} = V_n + \alpha \Vsearch,
\end{equation}
where we defined the search direction
\begin{equation}
    \Vsearch = V(\rho(V_n)) - V_n.
\end{equation}
The corresponding energy functional minimized by this fixed-point problem is
\begin{equation}
    \label{eqn:energyV}
    \I(V) = \E(\ffd(H_0 + \diagm(V))).
\end{equation}
Compared to an algorithm based on Kohn-Sham Hamiltonians
as suggested in Section \ref{sec:hammix}
this formulation has the advantage that only vector-sized potentials $V_n$
instead of matrix-sized quantities need to be handled.

The analysis of the previous sections carries forward straightforwardly
to the potential mixing setting.
In particular one identifies as the analogue of $\bm{K}$
the Hessian of $g$, i.e.~the (two-point) Hartree-exchange-correlation kernel $K$,
and as the analogue of $\bm{\chi_0}$
the derivative of $V(\rho)$, which is the independent-particle susceptibility $\chi_0$.
The latter becomes apparent by comparing \eqref{eqn:chi4}
to the Adler-Wiser formula for $\chi_0$~\cite{Adler1962,Wiser1963}
\begin{equation}
\label{eqn:chi}
\begin{aligned}
  \chi_0(V)
    &= \sum_{i=1}^{N_b} \sum_{j=1}^{N_b}
    \frac{\ffd(\varepsilon_i) - \ffd(\varepsilon_j)}{\varepsilon_i - \varepsilon_j}      |\phi_i^\ast \phi_j\rangle
      \langle \phi_i^{*} \phi_j|
\end{aligned}
\end{equation}
in which $(\varepsilon_i, \phi_i)$ denotes the eigenpairs of $H_0 + \diagm(V)$.
Both $K$ and $\chi_0$ arise naturally when considering the Jacobian matrix
\begin{equation}
    J_\alpha = 1 - \alpha \big(1 - K(V_\ast) \chi_0(V_\ast) \big)
    \label{eqn:Jacobian}
\end{equation}
of the potential-mixing SCF iteration \eqref{eqn:potmix} near a fixed point $V_\ast$.
If the eigenvalues of $J_\alpha$ are between $-1$ and $1$ the potential-mixing
SCF iterations converge.
By analogy with Hamiltonian mixing, Theorem \ref{thm:algorithm}
guarantees that global convergence can always be ensured by selecting $\alpha$ small enough.
In this respect our results from Section \ref{sec:hammix} strengthen
a number of previous results~\cite{dederichs1983self,gonze1996towards,cances2020convergence},
which established local convergence for sufficiently small $\alpha$.

\subsection{Improving the search direction $\dVn$: Preconditioning and acceleration}
The Jacobian matrix \eqref{eqn:Jacobian} involves the dielectric
matrix $\epsilon(V) = 1 - K(V) \chi_0(V)$, which can become
badly conditioned for many systems. In such cases, a very small step
must be employed to ensure stability (smallest eigenvalue of
$J_{\alpha}$ larger than $-1$), which slows down convergence (largest
eigenvalue of $J_{\alpha}$ very close to $1$) to a level too slow to
be practical. A solution is to improve the search direction $\Vsearch$
to ensure faster convergence~\cite{Woods2019}. This is usually
achieved by a combination of techniques jointly referred to as
``mixing'', which amend $\Vsearch$ using both preconditioning as
well as convergence acceleration.

Employing a preconditioned search direction
\begin{equation}
    \Vsearch = P^{-1} [V(\rho(V_n)) - V_n]
    \label{eqn:search_precon}
\end{equation}
in a damped SCF iteration, the corresponding Jacobian becomes
\begin{equation}
    J_\alpha = 1 - \alpha P^{-1} \epsilon(V).
\end{equation}
Provided that the inverse $P^{-1}$ approximates the inverse dielectric
matrix $\epsilon^{-1}$ sufficiently well, the spectrum of
$P^{-1} \epsilon$ is close to 1, so that a larger damping $\alpha$ and
and faster iteration is possible. While suitable cheap preconditioners
$P$ are not yet known for all sources of bad conditioning in SCF
iterations, a number of successful strategies have been suggested.
Examples include Kerker mixing~\cite{Kerker1981} to improve SCF
convergence in metals or LDOS-based mixing~\cite{ldos} to tackle
heterogeneous metal-vacuum or metal-insulator systems. For a more
detailed discussion on this matter we refer the reader to
Reference~\onlinecite{ldos}.

An additional possibility to speed up convergence is to use black-box
convergence accelerators. These techniques
build up a history of the previous iterates $V_1, \ldots, V_n$
as well as the previous preconditioned residuals $P^{-1} R_1, \ldots, P^{-1} R_n$
(with $R_n = V(\rho(V_n)) - V_n$) and use this information to obtain the
next search direction $\Vsearch$.
The most frequently used acceleration technique in this context is variously
known as Pulay/DIIS/Anderson mixing/acceleration,
which we will refer to as Anderson acceleration.
This method obtains the search direction as a linear combination
\begin{equation}
    \begin{aligned}
    \Vsearch &= P^{-1} R_{n} \\
    &\hspace{1.3em}+ \frac 1 \alpha \sum_{i=1}^{n-1} \beta_{i}\big(V_{i} + \alpha P^{-1} R_{i} - V_{n} - \alpha P^{-1} R_{n}\big)
    \end{aligned}
    \label{eqn:Anderson}
\end{equation}
where the expansion coefficients $\beta_{i}$ are found by minimizing
\begin{equation}
    \norm{P^{-1} R_n + \sum_{i=1}^{n-1} \beta_i \left(P^{-1} R_i - P^{-1} R_n\right)}.
    \label{eqn:AndersonMinimization}
\end{equation}
In practice, it is impractical to keep a potentially large number of
past iterates, and only the last 10 iterates are taken into
account. Furthermore, the associated linear least squares problem can become
ill-conditioned \cite{Walker2011}. We use the simple strategy of
discarding past iterates to ensure a maximal conditioning of $10^{6}$.

This method is known to be equivalent to a multisecant Broyden method.
In the linear regime and with infinite history Anderson acceleration
is further equivalent to the well-known GMRES method to solve linear
equations. For details see Reference~\cite{Chupin2020} and References therein.
Provided that nonlinear effects are
negligible, Anderson acceleration typically inherits the favorable
convergence properties of Krylov methods~\cite{Saad2003}, explaining
their frequent use in the DFT context. However, especially at the
beginning of the SCF iterations or when treating systems that feature
many close SCF minima, nonlinear effects can become important. In
such cases the behavior of Anderson is more complex and
mathematically not yet fully understood. In particular the dependence
of the convergence behavior on numerical parameters such as the
chosen damping can become less regular and harder to interpret, as we
will see in our numerical examples in Sections \ref{sec:nonlinear} and
\ref{sec:tm}.

\section{Adaptive damping algorithm}
\label{sec:adaptive}
\newcommand{\atrial}{\widetilde{\alpha}}
\newcommand{\atmin}{\atrial_\text{min}}

Up to now we have assumed that the step size $\alpha$ is constant,
reflecting common practice in plane-wave DFT computations. We now describe the
main contribution of this paper, an algorithm to adapt this step size
to increase robustness and minimize user intervention into the
convergence process. At step $n$ of the algorithm, given a trial
potential $V_{n}$, we compute the search direction $\delta V_{n}$
through \eqref{eqn:Anderson}, and look for a step $\alpha_{n}$ to take as
\begin{align}
  V_{n+1} = V_{n} + \alpha_{n} \delta V_{n}
\end{align}
Note that the definition of $\delta V_{n}$ itself in
\eqref{eqn:Anderson} depends on a stepsize $\alpha$; since our scheme
will adapt $\alpha_{n}$ to $\delta V_{n}$, we cannot just take
$\alpha=\alpha_{n}$ in \eqref{eqn:Anderson}, and so we use for
$\alpha$ a trial damping $\atrial$ (to be discussed in Section
\ref{sec:atrial}).

To select $\alpha_{n}$, we could try to minimize $\I(V_{n+1})$, or
employ an Armijo line-search strategy. However,
each evaluation of $\I$ is very costly, and it is therefore desirable
to obtain efficient approximate schemes. The energy
$\I(V + \alpha \dVn)$, can be expanded as
\begin{equation}
\begin{aligned}
    \I(V_n + \alpha \dVn)
    &= \I(V_n) + \alpha \langle \chi_0(V_n) R_{n}, \dVn\rangle\\
               &+ \frac12 \alpha^2 \langle \dVn , d^2 \I(V_n) \cdot \dVn \rangle
               + O(\alpha^3 \|\dVn\|^{3}),
\end{aligned}
\label{eq:energy_expansion}
\end{equation}
where we have used $\nabla \I(V_n) = \chi_{0}(V_{n}) R_{n}$. This
approximation is good for small dampings $\alpha$ and/or close
to the solution, when $\dVn$ is small.
The object $d^{2} \I$ is complicated and expensive to compute in
general. However, close to a fixed point, we can use the expression
\eqref{eqn:hessI} to write $d^{2} \I(V_{n}) \approx
\chi_{0}(V_{n})(1-K(V_n)) \chi_0(V_n)$. We can then approximate the terms in \eqref{eq:energy_expansion},
leading to the model
\newcommand{\model}{\varphi_n}
\begin{equation}
    \begin{aligned}
    \model(\alpha) &= \I(V_n)
    + \alpha \left\langle R_{n}, \chi_0(V_n) \dVn \right\rangle\\
    &- \frac12 \alpha^2 \left\langle \chi_0(V_n) \dVn, 
      \big[1-K(V_n) \chi_0(V_n)\big] \dVn \right\rangle
    \end{aligned}
    \label{eqn:modelFirst}
\end{equation}
for the energy, where we have used the self-adjointness of
$\chi_0(V_n)$ to make it act only on $\dVn$. To compute the coefficients in this model, we still need to
compute $\chi_{0}(V_{n}) \dVn$, a costly operation. However, for all
$\alpha$ we have to first order
\begin{align}
  \alpha \chi_{0}(V_{n}) \dVn = \rho(V_{n}+\alpha \dVn) - \rho(V_{n}) + O(\alpha^{2}\|\dVn\|^{2}).
  \label{eqn:model_coeffs}
\end{align}
Note that if we set $V_{n+1} = V_{n} + \alpha_{n} \dVn$ and then proceed along
the iterative algorithm, we will have to compute $\rho(V_{n+1})$ in any case.
An approximation to the coefficients of the model $\model$ can therefore
be constructed without any extra diagonalization.

This is the basis of the adaptive damping scheme described in Algorithm \ref{alg:adaptive}.
Since $\rho(V_{n})$ is already known (it is needed to construct $\dVn$),
the only expensive step in this algorithm is the computation of $\rho(V_{n+1})$,
which occurs only once per loop iteration.
In particular, when set to always accept $V_{n+1}$, this algorithm reduces to the
standard damped SCF algorithm. Notice that the algorithm only allows $\alpha_n$ to shrink between iterations.
As a result (i) the model $\model$ provides better and better damping predictions
and (ii) keeping in mind our analysis of Section \ref{sec:hammix}
the proposed tentative steps $V_{n+1}$ become more likely to be accepted.

\begin{algorithm}[H]
    \caption{Adaptive damping algorithm}
    \label{alg:adaptive}
    \begin{algorithmic}[1]
    \Require{
        Current iterate $V_{n}$, search direction $\Vsearch$, trial damping $\atrial$
    }
    \Ensure{Damping $\alpha_{n}$, next iterate $V_{n+1}$}
    \State $\alpha_n \gets \atrial$
    \Loop
        \State Make tentative step $V_{n+1} = V_n + \alpha_n \Vsearch$
        \State Compute $\rho(V_{n+1}), \I(V_{n+1})$ (the expensive step)
        \If{accept $V_{n+1}$ (see Section \ref{sec:step_acceptance})}
            \State \textbf{break}
        \Else
        \State Build the coefficients of the model $\model$
            \If{model $\model$ is good (see Section \ref{sec:quality_model})}
                \State $\alpha_n \gets \argmin_\alpha \model(\alpha)$
                \State Scale $\alpha_n$ to ensure $\abs{\alpha_n}$ is strictly decreasing
            \Else
                \State $\alpha_n \gets \frac{\alpha_n}{2}$
            \EndIf
        \EndIf
    \EndLoop
    \end{algorithmic}
  \end{algorithm}

We complete the description of the algorithm by specifying some practical points:
when to accept a step,
how to determine whether a model is good, how to select the initial
trial step $\widetilde \alpha$ and how to integrate adaptive damping
with Anderson acceleration.

\subsection{Step acceptance}
\label{sec:step_acceptance}
We accept the step as soon as the proposed next iterate
$V_{n+1} = V_n + \alpha_n \Vsearch$ satisfies
\begin{equation}
    \begin{aligned}
        \I(V_{n+1}) &< \I(V_n)
        &\text{or}\quad
        \norm{P^{-1} R_{n+1}} &< \norm{P^{-1} R_n},
    \end{aligned}
    \label{eqn:acceptance}
\end{equation}
i.e. if either the energy or the preconditioned residual decreases.
Although accepting steps higher in energy may decrease the robustness
of the algorithm, we found in practice that accepting steps that
decrease the residual helps keeping the method effective in the later
stages of convergence, when the Anderson acceleration is able to take
efficient steps that may slightly increase the energy but are not
worth reverting.

\subsection{Quality of the model $\model$}
\label{sec:quality_model}
Our model $\model$ makes various assumptions that might not hold in
practice, especially far from convergence. However, by comparing
the actual energy $\I(V_n+\alpha_{n}\dVn)$ to the prediction $\model(\alpha_{n})$, we can
inexpensively check the quality of the model. We do this by computing the ratio
\begin{equation}
  r_n = \frac{\abs{\I(V_n + \alpha_{n} \dVn) - \model(\alpha_{n})}}{\abs{\I(V_n + \alpha_{n} \dVn) - \I(V_n)}},
  \label{eqn:relerrormodel}
\end{equation}
which should be small if the model is accurate.
We deem the model good enough if
\begin{equation}
    r_n < 0.1 \qquad \text{and} \qquad \text{$\model$ has a minimum}.
    \label{eqn:modelacceptance}
\end{equation}
Notice that the minimizer of $\model$ may not necessarily be positive.
For particularly accurate models ($r_n < 0.01$) we additionally
allow backward steps (i.e.~$\alpha_n < 0$),
which turned out to overall improve convergence in our tests.

\subsection{Choice of the trial step $\widetilde \alpha$}
\label{sec:atrial}
To ensure that as many as possible SCF steps only require a single line search step,
we dynamically adjust $\atrial$ between two subsequent SCF steps.
A natural approach is to reuse the adaptively determined damping $\alpha_n$
as the $\atrial$ in the next line search,
which effectively shrinks $\atrial$ between SCF steps.
However, the algorithm may need small values of $\atrial$ in the
initial stages of convergence, and keeping these small values for too
long limits the eventual convergence rate.
To counteract the decreasing trend, we allow $\atrial$ to increase
if a line search was immediately successful (i.e.~$\alpha_n = \atrial$).
In this case we again use the model $\model$. If it is sufficiently
good (as described in Section \ref{sec:quality_model}), we set
\begin{equation}
    \atrial \gets \max\Big(\atrial, \ 1.1 \cdot \argmin_\alpha \model(\alpha) \Big).
\end{equation}
Otherwise $\atrial$ is left unchanged.

As an additional measure, to prevent the SCF from stagnating we enforce
$\atrial$ to not undershoot a minimal trial damping $\atmin$. We used
mostly $\atmin = 0.2$ as a baseline, and report varying this parameter
in the numerical experiments.

With this dynamic adjustment of $\atrial$, we checked that its initial
value $\atrial_0$, i.e.~the value used in the first SCF step, has
little influence on the overall convergence behavior. However,
in well-behaved cases, too small values for this parameter
lead to an unnecessary slowdown of the first few SCF steps. We therefore settled on $\atrial_0 = 0.8$ similar to standard recommendations
for the default damping~\cite{Kresse1996}.

\section{Numerical tests}
\label{sec:tests}
\label{sec:implementation}
The adaptive damping algorithm described in Section \ref{sec:adaptive}
was compared against a conventional preconditioned damped potential-mixing SCF scheme
featuring only a fixed damping.
For this we employed three kinds of test problems.
The first are calculations on aluminium systems of various size
including cases with an unsuitable
computational setup, i.e.~where charge sloshing is not prevented
by employing the Kerker preconditioner.
These are discussed in more detail in Section~\ref{sec:aluminium}.
The second, discussed in Section~\ref{sec:nonlinear},
is a gallium arsenide system which we previously found to feature
strongly nonlinear behavior in the initial SCF steps~\cite{ldos}.
Lastly in section Section~\ref{sec:tm} we will consider
Heusler systems and other transition-metal compounds,
which are generally found to be difficult to converge.

\begin{table*}
    \centering
\begin{tabular}{%
    l@{\extracolsep{1em}}l%
    @{\extracolsep{1em}}c@{\extracolsep{1em}}c%
    *{9}{@{\extracolsep{0.5em}}c}%
    @{\extracolsep{1em}}c@{\extracolsep{1em}}c%
}
\hline \hline
System&Precond.&&\multicolumn{10}{c}{fixed damping $\alpha$}&&adaptive\\
&&& 0.1 & 0.2 & 0.3 & 0.4 & 0.5 & 0.6 & 0.7 & 0.8 & 0.9 & 1.0 &&damping\\
\hline
\ce{Al8} supercell  &  Kerker\textsuperscript{a}  &&  $\times$ & 58 & 37 & 27 & 21 & 16 & 13 & \textbf{11} & 12 & 18 && 17 \\
\ce{Al8} supercell  &  None\textsuperscript{a}  &&  $\times$ & \textbf{52} & $\times$ & $\times$ & $\times$ & $\times$ & $\times$ & $\times$ & $\times$ & $\times$ && 24 \\
\ce{Al40} supercell  &  Kerker  &&  19 & 15 & 14 & 12 & \textbf{11} & 12 & 12 & 12 & 12 & 12 && 12 \\
\ce{Al40} supercell  &  None  &&  \textbf{38} & 40 & 40 & 39 & 44 & 50 & 49 & $\times$ & 76 & $\times$ && 44 \\
\ce{Al40} surface  &  Kerker  &&  $\times$ & $\times$ & $\times$ & $\times$ & $\times$ & $\times$ & $\times$ & $\times$ & $\times$ & $\times$ && $\times$ \\
\ce{Al40} surface  &  None  &&  \textbf{46} & 48 & 50 & 49 & 51 & 60 & 61 & 66 & 89 & $\times$ && 49 \\
\hline
\ce{Ga20As20} supercell  &  None  &&  \textbf{26} & 33 & 40 & 42 & 45 & 44 & 70 & 70 & 65 & 76 && 26 \\
\hline
\ce{CoFeMnGa}  &  Kerker  &&  $\times$ & $\times$ & $\times$ & $\times$ & 28 & \textbf{21} & 24 & 28 & 22 & 22 && 30 \\
\ce{Fe2CrGa}  &  Kerker  &&  $\times$ & $\times$ & $\times$ & 27 & $\times$ & $\times$ & \textbf{19} & 25 & $\times$ & 22 && 39 \\
\ce{Fe2MnAl}  &  Kerker  &&  $\times$ & 48 & $\times$ & $\times$ & $\times$ & 20 & 21 & 17 & 16 & \textbf{15} && 34 \\
\ce{FeNiF6}  &  Kerker  &&  $\times$ & $\times$ & $\times$ & $\times$ & $\times$ & $\times$ & $\times$ & 23 & 22 & \textbf{21} && 24 \\
\ce{Mn2RuGa}  &  Kerker  &&  $\times$ & $\times$ & $\times$ & $\times$ & 37 & 24 & 23 & \textbf{22} & 23 & 23 && 36 \\
\ce{Mn3Si}  &  Kerker  &&  $\times$ & $\times$ & $\times$ & $\times$ & 26 & 30 & 22 & \textbf{20} & $\times$ & $\times$ && $\times$ \\
\ce{Mn3Si} (AFM)\textsuperscript{b}  &  Kerker  &&  $\times$ & $\times$ & 58 & 29 & 31 & 30 & \textbf{20} & 22 & 26 & 28 && 35 \\
\hline
\ce{Cr19} defect  &  Kerker  &&  $\times$ & $\times$ & $\times$ & 74 & 46 & 48 & 46 & \textbf{41} & 47 & 53 && 48 \\
\ce{Fe28W8} multilayer  &  Kerker  &&  \textbf{32} & 34 & 37 & 34 & 38 & 43 & 41 & 48 & $\times$ & $\times$ && 37 \\
\hline \hline
\end{tabular}
\begin{minipage}{0.64\textwidth}
    \vspace{0.15em}
    \begin{flushleft}
    \footnotesize
    \textsuperscript{a}without Anderson acceleration\\
    \textsuperscript{b}initial guess with antiferromagnetic spin ordering
    \end{flushleft}
\end{minipage}
    \caption{Number of Hamiltonian diagonalizations required to obtain convergence
        in the energy to $10^{-10}$ with a cross ($\times$)
        denoting a failure to converge within $100$ diagonalizations.
        Except where otherwise noted Anderson acceleration has been employed
        and for the transition-metal systems (third/fourth group of compounds)
        a ferromagnetic initial guess has been used.
        On supercells atomic positions were slightly randomised.
        Computational details are given in the text.
        Notice that the transition-metal systems
        may not converge to the same SCF solution for each calculation.
    }
    \label{tab:convtable}
\end{table*}

For our tests we used the implementation of the adaptive damping algorithm available
in the density-functional toolkit (DFTK)~\cite{DFTKjcon,DFTK},
a flexible open-source Julia package for plane-wave density-functional theory simulations.
For all calculations we used Perdew-Burke-Ernzerhof~(PBE) exchange-correlation
functional~\cite{Perdew1996} as implemented in the
libxc~\cite{Lehtola2018} library, and Goedecker-Teter-Hutter
pseudopotentials~\cite{Goedecker1996}.
Depending on the system, a kinetic energy cutoff between $20$ and $45$ Hartree
as well as an unshifted Monkhorst-Pack with a maximal $k$-point spacing of
at most $0.14$ inverse Bohrs was used.
For the Heusler systems this was reduced to at most $0.08$ inverse Bohrs.
With the exception of the gallium arsenide system
a Gaussian smearing scheme with width of 0.001 Hartree was employed.
For the systems containing transition-metal elements collinear spin polarization
was allowed and the initial guess was constructed assuming
ferromagnetic spin ordering except when otherwise noted.
Notice, that this initial guess is generally not close to the
final spin ordering, see Section \ref{sec:tm} for discussion regarding this choice.
The full computational details for each system (including the employed structures)
as well as instructions how to reproduce all results of this paper
can be found in our repository of supporting information~\cite{reproducers}.

Table~\ref{tab:convtable} summarizes the required number of
Hamiltonian diagonalizations to converge the SCF energy to an error of
$10^{-10}$ Hartree for various fixed dampings $\alpha$ as well as the
adaptive damping algorithm. We carefully verified the obtained solutions
to be stationary points by monitoring the SCF residual $R_n$.
Note that for the adaptive damping
procedure the number of SCF steps is not identical to the number of
Hamiltonian diagonalizations, since multiple tentative steps might be
required until a step is accepted. Since iterative diagonalization
overall dominates the cost of the SCF procedure, the number of
diagonalizations provides a better metric to compare between the cost
of both damping strategies.

\subsection{Inadequate preconditioning: Aluminum}
\label{sec:aluminium}
\begin{figure}
    \centering
    \includegraphics[width=0.48\textwidth]{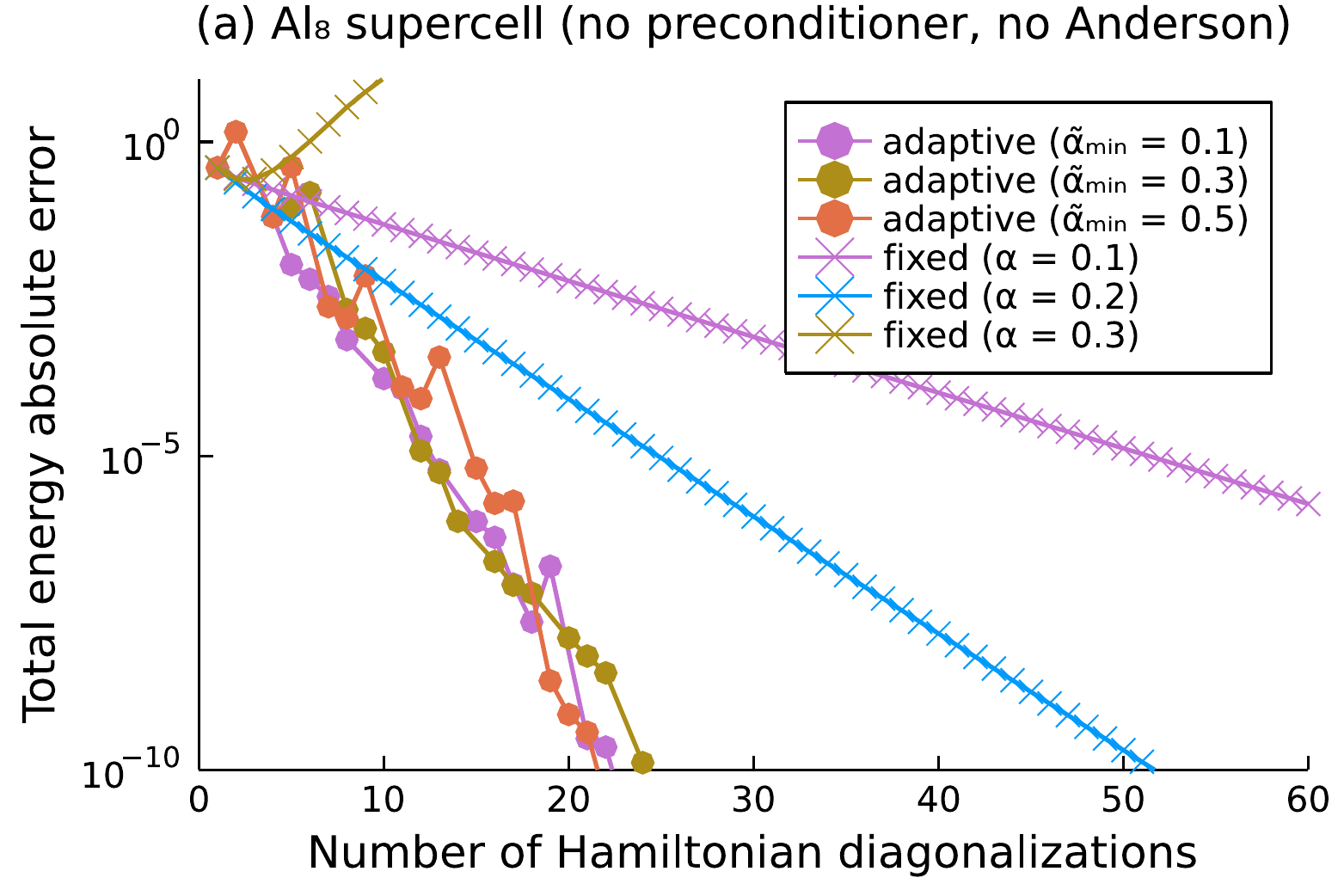}\\[0.5em]
    \includegraphics[width=0.48\textwidth]{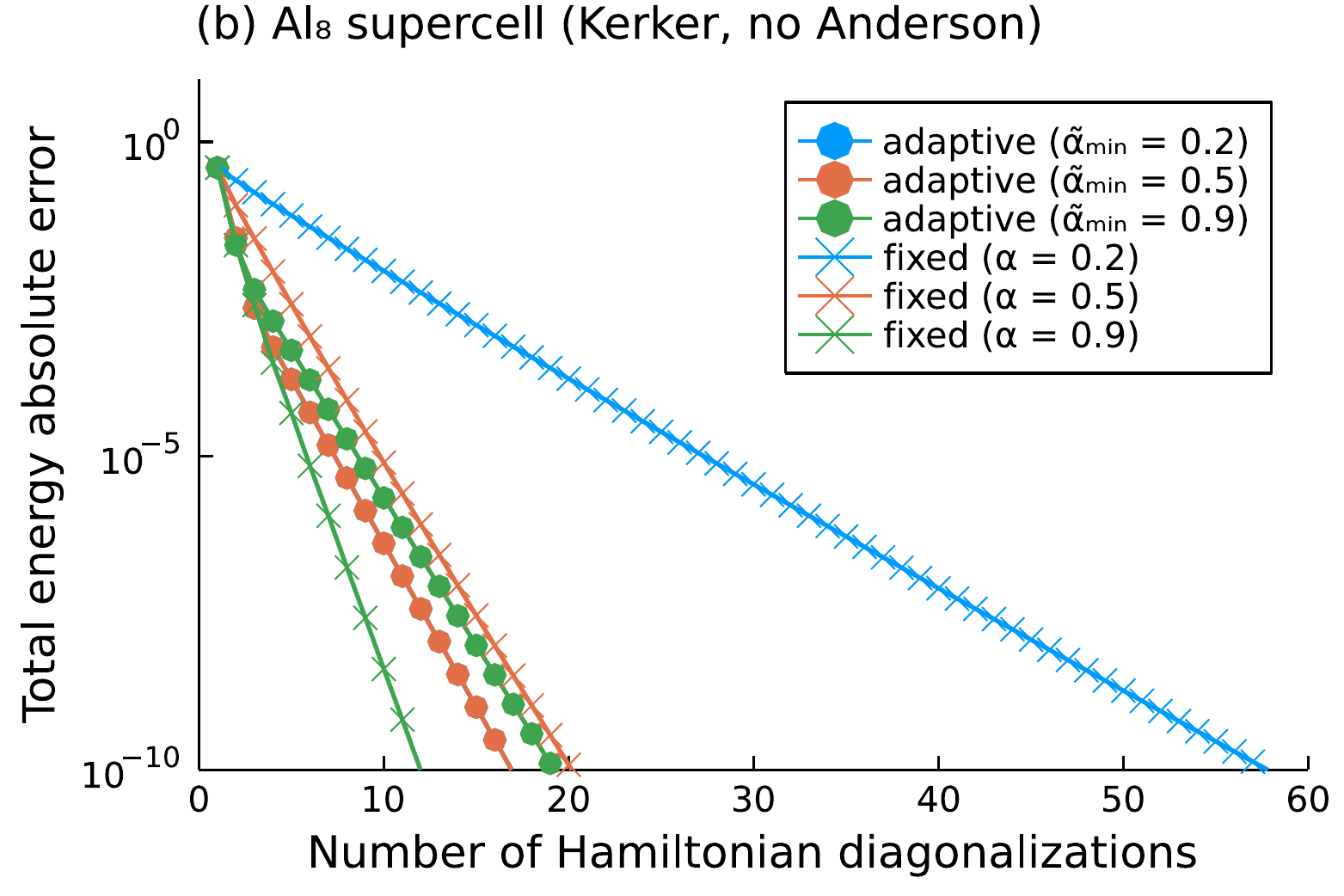}
    \caption{
        SCF convergence for a randomized aluminium supercell (8 atoms)
        without Anderson acceleration
        and using simple mixing (top) as well as Kerker mixing(bottom).
        The adaptive scheme converges robustly even in the unpreconditioned case,
        without requiring the manual selection of a step.
    }
    \label{fig:Al}
\end{figure}
To investigate the influence of the choice of a suboptimal preconditioner
on the convergence for both the fixed damping and adaptive damping strategies
we considered three aluminium test systems:
two elongated bulk supercells with 8 or 40 atoms
as well as a surface with 40 atoms
and a portion of vacuum of identical size.
For the elongated supercells both the initial guess as well as the atomic
positions were slightly rattled.

The results are summarized in the first segment of
Table~\ref{tab:convtable}. For the small \ce{Al8} system, where
Anderson acceleration was not used, representative convergence curves
are shown in Figure~\ref{fig:Al}. Due to the well-known
charge-sloshing behavior, SCF iterations on such metallic systems are
ill-conditioned. Without preconditioning small fixed damping values
$\alpha$ are thus required to obtain convergence, with only a small
window of damping values being able to achieve a convergence within
100 Hamiltonian applications. On the other hand in combination with
the matching Kerker preconditioning strategy~\cite{Kerker1981} large
fixed damping values generally converge more quickly.

In contrast the adaptive damping strategy is much less sensitive to
the choice of the minimal trial damping $\atmin$. Moreover, it leads
to a much improved convergence for the case without suitable
preconditioning while still maintaining similar costs if Kerker mixing
is employed.

\begin{figure}
    \centering
    \includegraphics[width=0.48\textwidth]{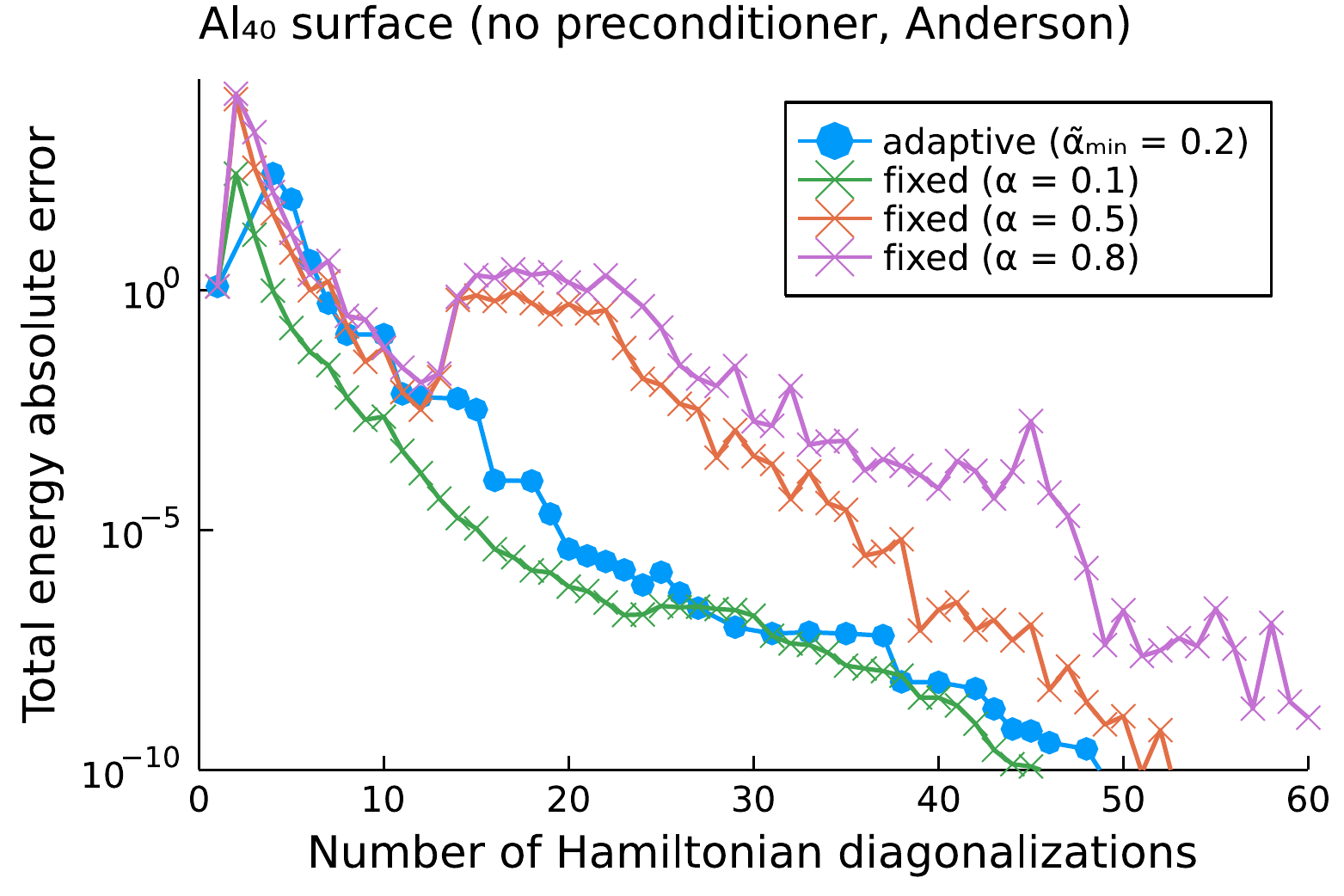}
    \caption{
        SCF convergence without preconditioning
        for an elongated supercell of an aluminium surface
        with 40 aluminium atom and a vacuum portion of equal size.
        A fixed step of $\alpha=0.1$ was optimal here,
        but the adaptive scheme gets very close
        performance without manual stepsize selection.
    }
    \label{fig:AlVac}
\end{figure}
These observations carry over to cases including Anderson acceleration
and larger aluminium systems, see Figure~\ref{fig:AlVac} for a
representative computation on an aluminium surface.
Notice that Kerker mixing is extremely badly suited for the
large aluminium surface, such that convergence is not obtained in 100 Hamiltonian
for any of the damping strategies,
see Reference \onlinecite{ldos} for a better preconditioner
in such inhomogeneous systems.
Overall employing adaptive damping therefore makes the reliability and
efficiency of the SCF less dependent on the choice of the
preconditioning strategy, while not requiring the user to manually
select a damping parameter.

\subsection{Strong nonlinear effects: Gallium arsenide}
\label{sec:nonlinear}

\begin{figure}
    \centering
    \includegraphics[width=0.48\textwidth]{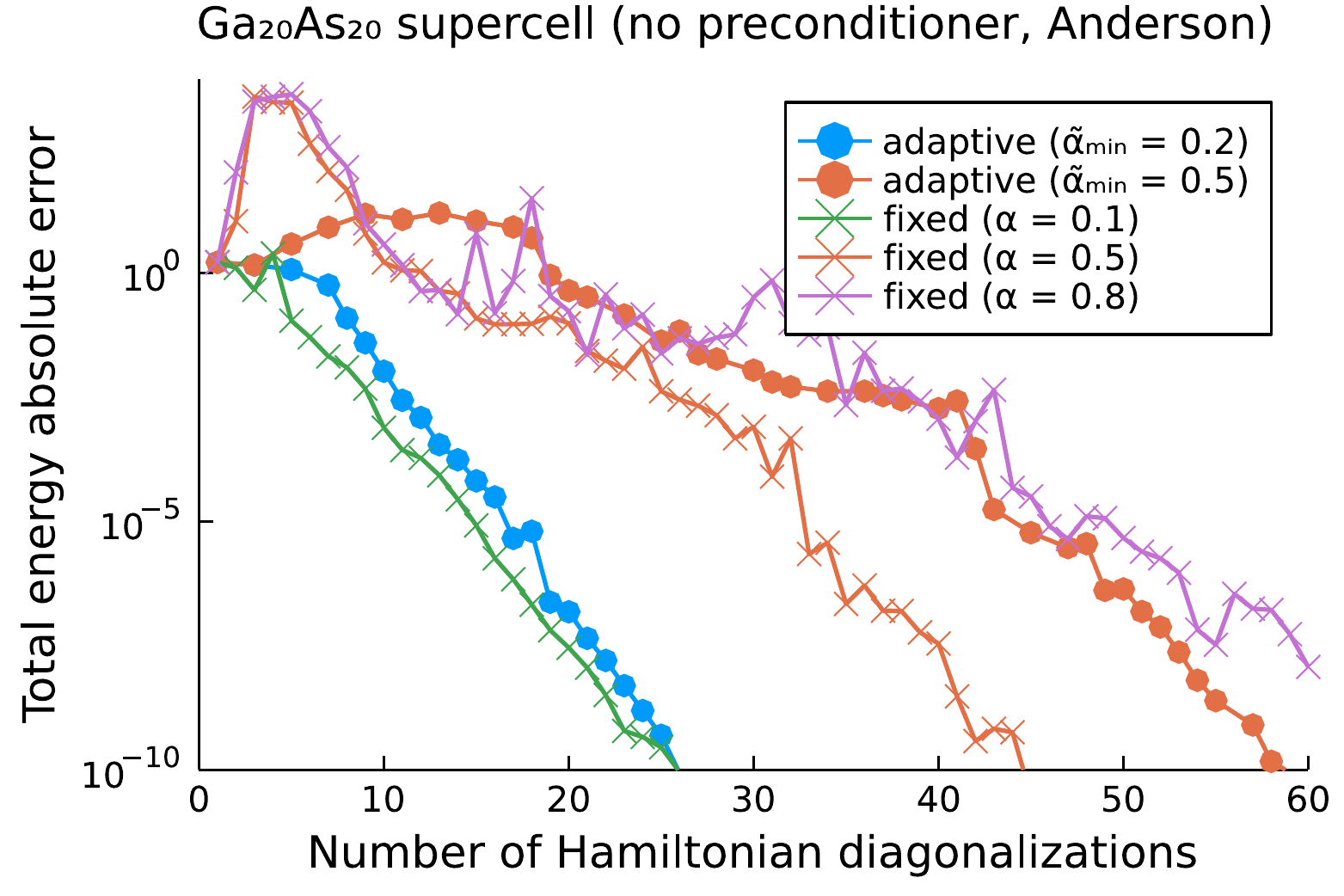}
    \caption{Elongated gallium arsenide supercell
        (20 gallium and 20 arsenide atoms)
    with slightly randomized atomic positions.
    For all cases, simple mixing and Anderson acceleration are
    employed. }
    \label{fig:GaAs}
\end{figure}
In previous work we identified elongated supercells of gallium arsenide
with slightly perturbed atomic positions to
be a simple system that still exhibits strong nonlinear effects
when the SCF is far from convergence~\cite{ldos}.
In the convergence profiles of these systems
this manifests by the error shooting up abruptly with Anderson failing
to quickly recover. In Figure~\ref{fig:GaAs}, for example, the error
increases steeply between Hamiltonian diagonalizations 3 and 6 for the
fixed-damping approaches with stepsizes beyond $0.1$. It should be
noted that this behavior is an artefact of the interplay of Anderson
acceleration and damped SCF iterations on these systems, which is not
observed in case Anderson acceleration is not employed. For more
details see the discussion of the gallium arsenide case in
Ref.~\onlinecite{ldos}.

For the calculations employing a fixed damping strategy only small
damping values of $\alpha=0.1$ are able to prevent this behavior.
Already slightly larger damping values noticably increase the number
of Hamiltonian diagonalizations required to reach convergence (compare
Table~\ref{tab:convtable}), and thus a careful selection of the
damping value is in order for such systems. In contrast the proposed
adaptive damping strategy with our baseline minimal trial damping of
$\atmin = 0.2$ automatically detects the unsuitable Anderson steps and
downscales them. As a result an optimal or near-optimal cost is
obtained without any manual parameter tuning. For comparison, we also
display in Figure~\ref{fig:GaAs} the results with a large value of
$\atmin=0.5$, which prevents the damping algorithm to avoid the
nonlinear effects.

\subsection{Challenging transition-metal compounds}
\label{sec:tm}

In this section we discuss two types of transition-metal systems.
First, we consider a selection of smaller primitive unit cells,
including the mixed iron-nickel fluoride \ce{FeNiF6}
as well a number of Heusler-type alloy structures,
see the third group of Table \ref{tab:convtable}.
These structures were found found in the course of high-throughput computations
to be difficult to converge~\cite{tricky}.
Moving to larger systems we considered
an elongated chromium supercell with a single vacancy defect
as well as a layered iron-tungsten system,
see the fourth group of Table \ref{tab:convtable}.
Both test cases were taken from previous studies~\cite{Winkelmann2020,Marks2021}
on SCF algorithms.

\begin{figure}
    \centering
    \includegraphics[width=0.48\textwidth]{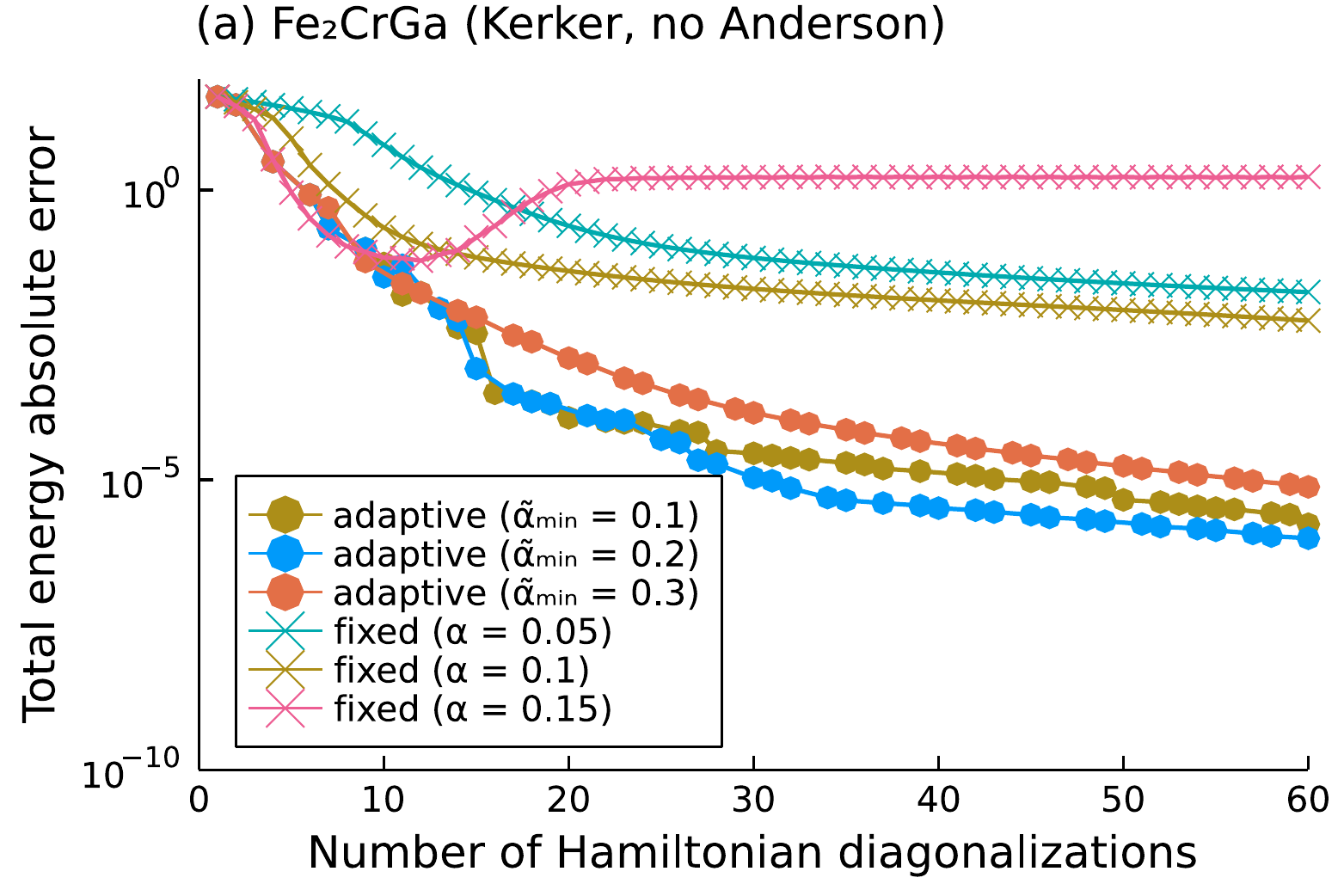}\\[0.5em]
    \includegraphics[width=0.48\textwidth]{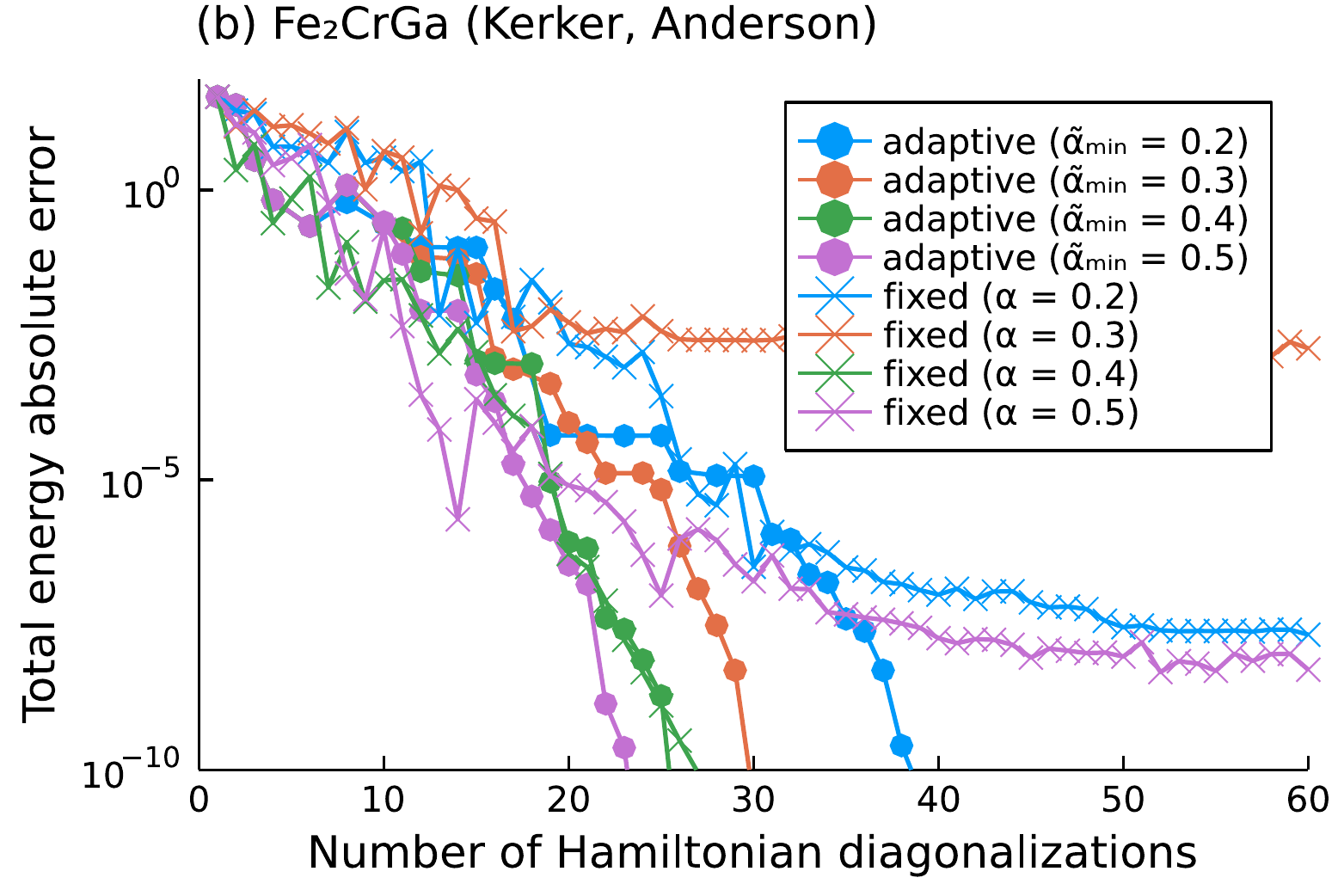}
    \caption{
        Convergence of the \ce{Fe2CrGa} Heusler alloy with Kerker mixing
        without Anderson acceleration (top)
        and
        with Anderson acceleration (bottom).
        Notice that the SCF calculations do not necessarily converge to
        the same local SCF minimum.
    }
    \label{fig:Fe2CrGa}
\end{figure}

In particular the Heusler
compounds are known to exhibit rich and unusual magnetic and electronic properties.
From our test set, for example, \ce{Fe2MnAl} shows halfmetallic behaviour,
i.e. a vanishing density of states at the Fermi level in only the minority
spin channel~\cite{Belkhouane2015}.
Other compounds, such as \ce{Mn2RuGa} or \ce{CoFeMnGa}
show an involved pattern of ferromagnetic and antiferromagnetic coupling
of the neighboring transition-metal sites~\cite{Wollmann2015,Shi2020}.
Such effects are closely linked to the $d$-orbitals forming localized states
near the Fermi level~\cite{He2018,Jiang2021}
and imply that there are a multiple accessible spin configurations,
which are close in energy.
Unfortunately these two properties also make Heusler compounds
difficult to converge using standard methods.
First, localized states near the Fermi level are a source of ill-conditioning for the
SCF fixed-point problem~\cite{ldos}, with no cheap and widely
applicable preconditioning strategy being available.
Second, the abundance of multiple spin configurations
implies a more involved SCF energy landscape with multiple SCF minima.
On such a landscape convergence may easily ``hesitate'' between
different local minima or stationary points.
Furthermore the setup of an appropriate initial guess, which ideally guides
the SCF towards the final spin ordering requires human expertise and is
hard to automatise in the high-throughput setting.
Albeit not fully appropriate for the systems we consider,
we followed the guess setup, which has also been used in the
aforementioned high-throughput procedure~\cite{tricky},
namely to start the calculations with an initial guess based on
ferromagnetic~(FM) spin ordering.

As a result in our tests,
calculations on Heusler systems without Anderson acceleration require
very small fixed damping values below $0.1$ even if the Kerker preconditioner is
used, see Figure~\ref{fig:Fe2CrGa} (a). The adaptive damping strategy
improves the convergence behavior and in agreement with our previous
results partially corrects for the mismatch in preconditioner and initial guess.
Still, convergence is extremely slow.

An acceptable convergence is only accessible in combination with
Anderson acceleration. %
However, the Anderson-accelerated
fixed-damping SCF is very susceptible to the chosen damping $\alpha$,
see Figure~\ref{fig:Fe2CrGa} (b). In particular the lowest-energy SCF
minimum is only found within 100 Hamiltonian diagonalizations for
$\alpha=0.4$, $\alpha=0.7$, $\alpha=0.8$ and $\alpha=1.0$. Other fixed
damping values initially converge, but then convergence stagnates and
the error only reduces very slowly beyond around $30$ diagonalizations.

We investigated the source of this pathological behavior by restarting
the iterations after stagnation.
This did not noticeably alter the behavior,
eliminating the possibility that the history of the iterates
within the Anderson acceleration scheme
somehow ``jam'' the SCF into stagnation
in the strongly nonlinear regime
--- as can happen for instance in nonlinear
conjugate gradient methods~\cite{hager2006survey}.
Another possibility is that the iterations
somehow got into a particularly rough region of SCF energy landscape
between multiple stationary points, which is simply hard to escape.
However, this is not the case either. For instance on the \ce{Fe2CrGa}
system with a fixed damping of $0.3$, the restarted iterations did
converge quickly using an Anderson scheme with a small maximal
conditioning of $10^{2}$ for the linear least squares problem.
It would therefore appear that this phenomenon is due to inadequate
regularization of the least squares problem. We expect more
sophisticated techniques for controlling the Anderson
history~\cite{Chupin2020} to be worth investigating for such systems
in the future.

Because of this stagnation issue we found Anderson-accelerated SCF
iterations to become unreliable for our transition-metal test systems:
for fixed damping values below $\alpha=0.5$, hardly any calculation
converges. Notably, due to the non-trivial interplay with the Anderson
scheme, this result is the exact opposite to our theoretical
developments on damped SCF iterations in Section \ref{sec:hammix},
which suggested to reduce the damping to achieve reliable convergence.

Overall the transition-metal cases emphasize the difficulty in manually choosing
an appropriate fixed damping.
For a number of cases the window of converging damping values
is rather narrow, e.g.~consider \ce{Fe2CrGa}, \ce{FeNiF6}
or \ce{Mn3Si} (with the FM guess)
and in our tests only a single damping value of $\alpha=0.8$ fortitiously
manages to converge all systems.

\begin{figure}
    \centering
    \includegraphics[width=0.48\textwidth]{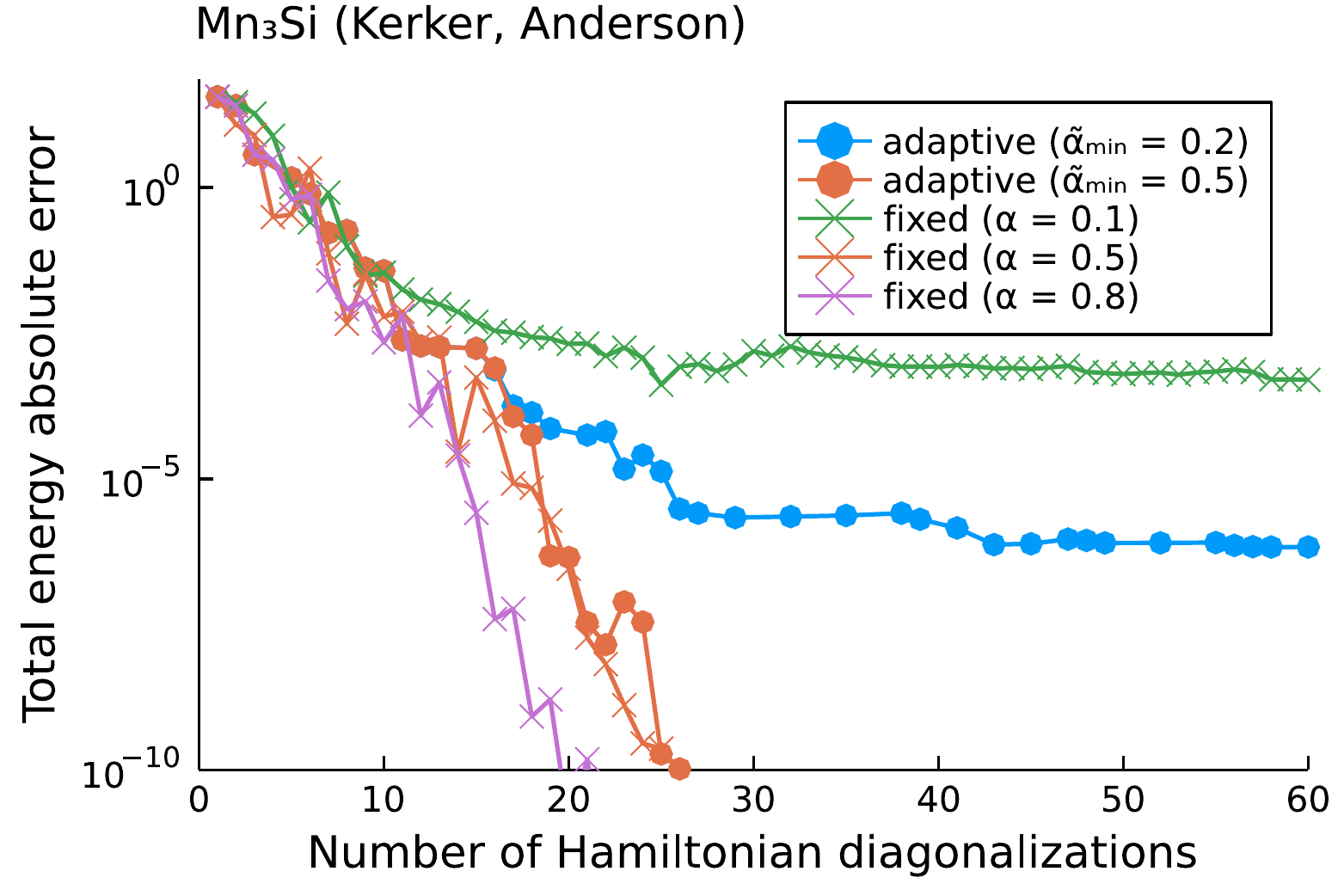}
    \caption{Convergence of the \ce{Mn3Si} Heusler compound
        with Kerker mixing and Anderson acceleration
        and starting from a ferromagnetically ordered initial guess.
        Small dampings are susceptible to stagnation
        induced by Anderson instabilities.
    }
    \label{fig:Mn3Si}
\end{figure}
In contrast the proposed adaptive damping strategy with our baseline
value of $\atmin = 0.2$ is less susceptible to the stagnation issue.
Across the unit cells and extended transition-metal systems we considered
we observed only a convergence failure in one test case, namely the
\ce{Mn3Si} Heusler alloy with the FM guess, see Figure \ref{fig:Mn3Si}.
For some test cases,
adaptive damping did cause a noticeable computational overhead: in
extreme cases (such as \ce{Fe2CrGa} or \ce{Fe2MnAl}) the number of
required Hamiltonian diagonalizations almost doubles. However, it
should be emphasized that no user adjustments were needed to obtain
these results, even though adaptive damping has been constructed on
the here invalid principle that smaller damping increases reliability.

Yet even on cases where Anderson instabilities cause non-convergence
of the adaptive scheme, fine-tuning is possible. Increasing the
minimal trial damping from $\atmin=0.2$ to $\atmin=0.5$, for example,
increases the minimal step size and thus lowers the risk of Anderson
stagnation. For all transition-metal cases we considered $\atmin=0.5$
strictly reduces the number of diagonalizations required to reach
convergence compared to $\atmin=0.2$, see for example
Figure~\ref{fig:Fe2CrGa} (b). Moreover this parameter adjustment even
resolves the convergence issues of \ce{Mn3Si}, see Figure~\ref{fig:Mn3Si}.
If manual intervention is possible
another option is to incorporating prior knowledge of the
final ground state electronic structure into the initial guess.
For the \ce{Mn3Si} case, for example, an improved initial guess based on
an antiferromagnetic spin ordering (AFM) between adjacent manganese layers
simplifies the SCF problem, such that both a larger range of fixed
damping values as well as the adaptive damping strategy give rise to
converging calculations.

\section{Conclusion}
\label{sec:conclusion}

We proposed a new linesearch strategy for SCF computations, based
on an efficient approximate model for the energy as a function of the
damping. Our algorithm follows four general principles: (a) the
algorithm should need no manual intervention from the user; (b) it
should be combinable with known effective mixing techniques such as
preconditioning and Anderson acceleration; (c) in ``easy'' cases where
convergence with a fixed damping is satisfactory it should not slow
down too much; and (d) it should be possible to relate it to schemes
with proved convergence guarantees. We demonstrated that our proposed
scheme fulfills all these objectives. With our default parameter choice
of $\atmin = 0.2$ the resulting adaptively damped SCF algorithm
is able to converge all of the ``easy'' cases faster or almost as fast as
the fixed-damping method with the best damping.
Simultaneously it is more robust than the fixed-damping method
on the ``hard'' cases we considered,
such as elongated bulk metals and metal surfaces without proper preconditioning
or Heusler-type transition-metal alloys.
In particular the latter kind of systems feature a very irregular convergence
behavior with respect to the damping parameter, making a robust manual damping
selection very challenging.
In practice the classification between ``easy'' and ``hard'' cases may well depend
on the considered system and the details of the computational setup,
e.g.~the employed mixing and acceleration techniques.
However, our scheme makes no assumptions about the details
how a proposed SCF step has been obtained.
We therefore believe adaptive damping to be a black-box stabilisation technique
for SCF iterations, which applies beyond the Anderson-accelerated setting
we have considered here.

Still, our results on these ``hard'' cases also highlight poorly-understood
limitations of the commonly used Anderson acceleration process. For example,
despite following standard recommendations to increase Anderson robustness,
we frequently observe SCF iterations to stagnate.
A more thorough understanding of this effect would
be an interesting direction for future research.

Our scheme was applied to semilocal density functionals in a
plane-wave basis set. It is not specific to plane-wave basis sets, and
we expect it to be similarly efficient in other ``large'' basis sets
frequently used in condensed-matter physics. For atom-centered basis sets,
like those common in quantum chemistry, direct mixing of the
density matrix is feasible, and likely more efficient. Our scheme does
not apply directly to hybrid functionals, where orbitals or Fock
matrices have to be mixed also; an extension to this case would be an
interesting direction for future research.

\section*{Acknowledgements}
This project has received funding from
the European Research Council (ERC) under
the European Union's Horizon 2020 research and innovation program
(grant agreement No 810367).
We are grateful to Marnik Bercx and Nicola Marzari
for pointing us to the challenging transition-metal structures
that stimulated most of the presented developments.
Fruitful discussions with Eric Cancès, Xavier Gonze and Benjamin Stamm
and provided computational time at Ecole des Points and RWTH Aachen University
are gratefully acknowledged.

\section*{Appendix: mathematical proofs}
\label{sec:proofs}
\noindent
\begin{lemma}
  \label{sec:lemma}
  Let
  \begin{align}
    H_{*} = \sum_{i=1}^{N_{\rm b}} \varepsilon_{i} |\phi_{i}\rangle\langle \phi_{i}|
  \end{align}
  with orthonormal $\phi_{i}$ and non-decreasing $\varepsilon_{i}$. Let
  $f : \R \to \R$ be a real analytic function in a neighborhood of $[\varepsilon_{1},\varepsilon_{N_{\rm b}}]$. Then the
  map $H \mapsto f(H)$ is analytic in a neighborhood of $H_{*}$, and
  \begin{align}
    {\bf df}(H_{*}) \cdot \delta H = \sum_{i=1}^{N_{\rm b}}\sum_{j=1}^{N_{\rm b}} \frac{f(\varepsilon_{i})-f(\varepsilon_{j})}{\varepsilon_{i}-\varepsilon_{j}}\langle  \phi_{i}, \delta H \phi_{j} \rangle |\phi_{i}\rangle\langle \phi_{j}| 
  \end{align}
  with the convention that $\frac{f(\varepsilon_{i}) -
    f(\varepsilon_{i})}{\varepsilon_{i}-\varepsilon_{i}} = f'(\varepsilon_{i})$.
\end{lemma}
\noindent \textbf{Proof of Lemma 1.}\qquad
  This is a classical result, known as the Daleckii-Krein theorem in
  linear algebra; see for instance Higham~\cite{higham2008functions}. To keep
  this paper self-contained, we
  follow here the proof in Levitt~\cite{levitt2020screening} in the analytic
  case. Since $f$ is analytic on
  $[\varepsilon_{1},\varepsilon_{N_{\rm b}}]$, it is analytic in a
  complex neighborhood. Let $\mathcal C$ be a positively oriented
  contour enclosing $[\varepsilon_{1},\varepsilon_{N_{\rm b}}]$. Then,
  for $H$ close enough to $H_{*}$, we have
  \begin{align}
    f(H) = \frac{1}{2\pi i}\oint_{\mathcal C} f(z) \frac {1} {z-H} dz
  \end{align}
  and analyticity of $f$ follows. For $\delta H$ small enough,
  \begin{equation}
  \begin{aligned}
    &f(H_{*}+\delta H) \\
    &= \frac 1 {2\pi i} \oint_{\mathcal C} f(z) \frac {1} {z-H_{*}-\delta H} dz\\
    &\approx f(H_{*}) + \frac 1 {2\pi i} \oint_{\mathcal C} f(z) \frac 1 {z-H_{*}}\delta H \frac 1 {z-H_{*}}dz\\
    &= f(H_{*}) + \frac 1 {2\pi i} \oint_{\mathcal C} {}\sum_{i=1}^{N_{\rm b}}\sum_{j=1}^{N_{\rm b}}  \frac {f(z) \langle  \phi_{i}, \delta H \phi_{j} \rangle }{(z-\varepsilon_{i})(z-\varepsilon_{j})} |\phi_{i}\rangle\langle \phi_{j}|dz\\
    &= f(H_{*}) + \sum_{i=1}^{N_{\rm b}}\sum_{j=1}^{N_{\rm b}} \frac{f(\varepsilon_{i})-f(\varepsilon_{j})}{\varepsilon_{i}-\varepsilon_{j}}\langle  \phi_{i}, \delta H \phi_{j} \rangle |\phi_{i}\rangle\langle \phi_{j}|
  \end{aligned}
  \end{equation}
  where $\approx$ means
  up to terms of order $O\left(\|\delta H\|^{2}\right)$.
\qed

\noindent \textbf{Proof of Theorem 1.}\qquad
  If $\alpha_{0} \le 1$, $H_{n}$ belongs to the
  convex hull spanned by $H_{0}$ and $\{\HKS(\ffd(H)), H \in \mathcal
  H\}$. On this compact set $X$, $\ffd$, $\I$ and their derivatives are
  bounded. We have for all $H \in X$
  \begin{equation}
    \label{eq:expansion_alphasquare}
  \begin{aligned}
    &\I(H + \alpha (\HKS - H)) \\
    &= \I(H) - \alpha \langle  \bm \Omega^{-1} (  \HKS -H), (  \HKS -H) \rangle + O(\alpha^{2})\\
    &= \I(H) - \alpha \langle  \bm \Omega \nabla\I(H), \nabla\I(H) \rangle + O(\alpha^{2})
  \end{aligned}
  \end{equation}
  where in this expression the functions $\bm \Omega$ and $\HKS$ are
  evaluated at $\ffd(H)$, and the constant in the $O(\alpha^{2})$ term is uniform in $n$.
  It follows that for $\alpha_{0}$ small enough, there is $c > 0$ such that
  \begin{align}
    \label{eq:guaranteed_energy_decay}
    \I(H_{n+1}) \le \I(H_{n}) - \alpha c \|\nabla\I(H_{n})\|^{2},
  \end{align}
  and therefore \mbox{$\nabla\I(H_{n}) \to 0$}, so that
  \mbox{$\HKS(\ffd(H_{n})) - H_{n} \to 0$}.

  We now proceed as in \citet{levitt2012convergence}. Let
  $\I_{*} = \lim_{n \to \infty}\I(H_{n})$. The set
  $\Gamma = \{H \in X,$ \mbox{$\I(H) = \lim_{n \to \infty}\I(H_{n})\}$} is
  non-empty and compact. Furthermore, $d(H_{n}, \Gamma) \to 0$; if
  this was not the case, we could extract by compactness of $X$ a
  subsequence at finite distance of $\Gamma$ converging to a
  $H_{*} \in X$ satisfying $\I(H_{*}) = \lim_{n \to \infty}\I(H_{n})$,
  which would imply that $H_{*} \in \Gamma$, a contradiction.

  At every point $H$ of $\Gamma$, by analyticity there is a neighborhood of $H$ in
  $\mathcal H$ such that the \L{}ojasiewicz inequality
  \begin{align}
    |\I(H') - \I_{*}|^{1-\theta_{H}} \le \kappa_{H} \|\nabla\I(H')\|
  \end{align}
  holds for some constants $\theta_{H} \in (0,1/2]$, $\kappa_{H} > 0$ \cite{levitt2012convergence,lojasiewicz1965ensembles}.
  By compactness, we can extract a finite covering of these
  neighborhoods, and obtain a \L{}ojasiewicz inequality with universal
  constants $\theta \in (0,1/2], \kappa > 0$ in a neighborhood of
  $\Gamma$. Therefore, for $n$ large enough, using the concavity
  inequality $x^{\theta} \le y^{\theta} + \theta y^{\theta-1}(x-y)$ with
  $x = \I(H_{n+1}) - \I_{*}$, $y = \I(H_{n}) - \I_{*}$, we get
  \begin{equation}
  \begin{aligned}
    \|\nabla\I(H_{n})\|^{2} &\le \frac 1 {\alpha c}{\I(H_{n}) - \I(H_{n+1})}\\
    & \le \frac 1 {\theta \alpha c} (\I(H_{n}) - \I_{*})^{1-\theta}\\
        &\hspace{1cm}\cdot
        \Big[\left(\I(H_{n}) - \I_{*}\right)^{\theta} - (\I(H_{n+1})-\I_{*})^{\theta} \Big]\\
    &\le \frac \kappa {\theta \alpha c} \|\nabla\I(H_{n})\|\\
        &\hspace{1cm}\cdot
        \Big[(\I(H_{n}) - \I_{*})^{\theta} - (\I(H_{n+1})-\I_{*})^{\theta} \Big]\\
    \|\nabla\I(H_{n})\|    &\le \frac \kappa {\theta \alpha c} \Big[(\I(H_{n}) - \I_{*})^{\theta} - (\I(H_{n+1})-\I_{*})^{\theta} \Big]
  \end{aligned}
  \end{equation}
  It follows that $\|\nabla\I(H_{n})\| $ is summable, and therefore
  that $\|H_{n+1} - H_{n}\|$ is; this implies convergence of $H_{n}$ to
  some $H_{*}$.
  When $\theta=1/2$ (or, in light of \eqref{eqn:hess_both}, when
  $\bm{d^{2} \E}(\ffd(H_{*}))$ is positive definite), we can get exponential convergence \cite{levitt2012convergence}.
  \qed

Note that the bounds used in the proof of the above statement (for
  instance, on $\alpha_{0}$) are extremely pessimistic, since they
  rely on the fact that the set of possible $P$ is bounded, and
  therefore all density matrices of the form $\ffd(\HKS(P))$ have
  occupations bounded away from $0$ and $1$, which results in bounded
  derivatives for $\Tr(s(P))$. A more careful analysis is needed to
  obtain better bounds (for instance, bounds that are better behaved
  in the zero temperature limit).

\noindent \textbf{Proof of Theorem 2.}
From \eqref{eq:expansion_alphasquare} it is easily seen that the
linesearch process stops in a finite number of iterations, independent
on $n$. This ensures that there is $\alpha_{\rm min} > 0$ such that
$\alpha_{\rm min} \le \alpha_{n} \le \alpha_{\rm max}$. From
\eqref{eq:armijo} it follows that a similar inequality to
\eqref{eq:guaranteed_energy_decay} holds, and the rest of the proof
proceeds as in that of Theorem 1.

\bibliography{article}

\begin{thebibliography}{50}%
\makeatletter
\providecommand \@ifxundefined [1]{%
 \@ifx{#1\undefined}
}%
\providecommand \@ifnum [1]{%
 \ifnum #1\expandafter \@firstoftwo
 \else \expandafter \@secondoftwo
 \fi
}%
\providecommand \@ifx [1]{%
 \ifx #1\expandafter \@firstoftwo
 \else \expandafter \@secondoftwo
 \fi
}%
\providecommand \natexlab [1]{#1}%
\providecommand \enquote  [1]{``#1''}%
\providecommand \bibnamefont  [1]{#1}%
\providecommand \bibfnamefont [1]{#1}%
\providecommand \citenamefont [1]{#1}%
\providecommand \href@noop [0]{\@secondoftwo}%
\providecommand \href [0]{\begingroup \@sanitize@url \@href}%
\providecommand \@href[1]{\@@startlink{#1}\@@href}%
\providecommand \@@href[1]{\endgroup#1\@@endlink}%
\providecommand \@sanitize@url [0]{\catcode `\\12\catcode `\$12\catcode
  `\&12\catcode `\#12\catcode `\^12\catcode `\_12\catcode `\%12\relax}%
\providecommand \@@startlink[1]{}%
\providecommand \@@endlink[0]{}%
\providecommand \url  [0]{\begingroup\@sanitize@url \@url }%
\providecommand \@url [1]{\endgroup\@href {#1}{\urlprefix }}%
\providecommand \urlprefix  [0]{URL }%
\providecommand \Eprint [0]{\href }%
\providecommand \doibase [0]{http://dx.doi.org/}%
\providecommand \selectlanguage [0]{\@gobble}%
\providecommand \bibinfo  [0]{\@secondoftwo}%
\providecommand \bibfield  [0]{\@secondoftwo}%
\providecommand \translation [1]{[#1]}%
\providecommand \BibitemOpen [0]{}%
\providecommand \bibitemStop [0]{}%
\providecommand \bibitemNoStop [0]{.\EOS\space}%
\providecommand \EOS [0]{\spacefactor3000\relax}%
\providecommand \BibitemShut  [1]{\csname bibitem#1\endcsname}%
\let\auto@bib@innerbib\@empty
\bibitem [{\citenamefont {Woods}\ \emph {et~al.}(2019)\citenamefont {Woods},
  \citenamefont {Payne},\ and\ \citenamefont {Hasnip}}]{Woods2019}%
  \BibitemOpen
  \bibfield  {author} {\bibinfo {author} {\bibfnamefont {N.~D.}\ \bibnamefont
  {Woods}}, \bibinfo {author} {\bibfnamefont {M.~C.}\ \bibnamefont {Payne}}, \
  and\ \bibinfo {author} {\bibfnamefont {P.~J.}\ \bibnamefont {Hasnip}},\
  }\href {\doibase 10.1088/1361-648x/ab31c0} {\bibfield  {journal} {\bibinfo
  {journal} {Journal of Physics: Condensed Matter}\ }\textbf {\bibinfo {volume}
  {31}},\ \bibinfo {pages} {453001} (\bibinfo {year} {2019})}\BibitemShut
  {NoStop}%
\bibitem [{\citenamefont {Lehtola}\ \emph {et~al.}(2020)\citenamefont
  {Lehtola}, \citenamefont {Blockhuys},\ and\ \citenamefont
  {Van~Alsenoy}}]{Lehtola2020}%
  \BibitemOpen
  \bibfield  {author} {\bibinfo {author} {\bibfnamefont {S.}~\bibnamefont
  {Lehtola}}, \bibinfo {author} {\bibfnamefont {F.}~\bibnamefont {Blockhuys}},
  \ and\ \bibinfo {author} {\bibfnamefont {C.}~\bibnamefont {Van~Alsenoy}},\
  }\href {\doibase 10.3390/molecules25051218} {\bibfield  {journal} {\bibinfo
  {journal} {Molecules}\ }\textbf {\bibinfo {volume} {25}},\ \bibinfo {pages}
  {1218} (\bibinfo {year} {2020})}\BibitemShut {NoStop}%
\bibitem [{\citenamefont {Jain}\ \emph {et~al.}(2016)\citenamefont {Jain},
  \citenamefont {Shin},\ and\ \citenamefont {Persson}}]{Jain2016}%
  \BibitemOpen
  \bibfield  {author} {\bibinfo {author} {\bibfnamefont {A.}~\bibnamefont
  {Jain}}, \bibinfo {author} {\bibfnamefont {Y.}~\bibnamefont {Shin}}, \ and\
  \bibinfo {author} {\bibfnamefont {K.~A.}\ \bibnamefont {Persson}},\ }\href
  {\doibase 10.1038/natrevmats.2015.4} {\bibfield  {journal} {\bibinfo
  {journal} {Nature Reviews Materials}\ }\textbf {\bibinfo {volume} {1}}
  (\bibinfo {year} {2016}),\ 10.1038/natrevmats.2015.4}\BibitemShut {NoStop}%
\bibitem [{\citenamefont {Alberi}\ \emph {et~al.}(2019)\citenamefont {Alberi},
  \citenamefont {Nardelli}, \citenamefont {Zakutayev}, \citenamefont {Mitas},
  \citenamefont {Curtarolo}, \citenamefont {Jain}, \citenamefont {Fornari},
  \citenamefont {Marzari}, \citenamefont {Takeuchi}, \citenamefont {Green},
  \citenamefont {Kanatzidis}, \citenamefont {Toney}, \citenamefont {Butenko},
  \citenamefont {Meredig}, \citenamefont {Lany}, \citenamefont {Kattner},
  \citenamefont {Davydov}, \citenamefont {Toberer}, \citenamefont {Stevanovic},
  \citenamefont {Walsh}, \citenamefont {Park}, \citenamefont {Aspuru-Guzik},
  \citenamefont {Tabor}, \citenamefont {Nelson}, \citenamefont {Murphy},
  \citenamefont {Setlur}, \citenamefont {Gregoire}, \citenamefont {Li},
  \citenamefont {Xiao}, \citenamefont {Ludwig}, \citenamefont {Martin},
  \citenamefont {Rappe}, \citenamefont {Wei},\ and\ \citenamefont
  {Perkins}}]{Alberi2019}%
  \BibitemOpen
  \bibfield  {author} {\bibinfo {author} {\bibfnamefont {K.}~\bibnamefont
  {Alberi}}, \bibinfo {author} {\bibfnamefont {M.~B.}\ \bibnamefont
  {Nardelli}}, \bibinfo {author} {\bibfnamefont {A.}~\bibnamefont {Zakutayev}},
  \bibinfo {author} {\bibfnamefont {L.}~\bibnamefont {Mitas}}, \bibinfo
  {author} {\bibfnamefont {S.}~\bibnamefont {Curtarolo}}, \bibinfo {author}
  {\bibfnamefont {A.}~\bibnamefont {Jain}}, \bibinfo {author} {\bibfnamefont
  {M.}~\bibnamefont {Fornari}}, \bibinfo {author} {\bibfnamefont
  {N.}~\bibnamefont {Marzari}}, \bibinfo {author} {\bibfnamefont
  {I.}~\bibnamefont {Takeuchi}}, \bibinfo {author} {\bibfnamefont {M.~L.}\
  \bibnamefont {Green}}, \bibinfo {author} {\bibfnamefont {M.}~\bibnamefont
  {Kanatzidis}}, \bibinfo {author} {\bibfnamefont {M.~F.}\ \bibnamefont
  {Toney}}, \bibinfo {author} {\bibfnamefont {S.}~\bibnamefont {Butenko}},
  \bibinfo {author} {\bibfnamefont {B.}~\bibnamefont {Meredig}}, \bibinfo
  {author} {\bibfnamefont {S.}~\bibnamefont {Lany}}, \bibinfo {author}
  {\bibfnamefont {U.}~\bibnamefont {Kattner}}, \bibinfo {author} {\bibfnamefont
  {A.}~\bibnamefont {Davydov}}, \bibinfo {author} {\bibfnamefont {E.~S.}\
  \bibnamefont {Toberer}}, \bibinfo {author} {\bibfnamefont {V.}~\bibnamefont
  {Stevanovic}}, \bibinfo {author} {\bibfnamefont {A.}~\bibnamefont {Walsh}},
  \bibinfo {author} {\bibfnamefont {N.-G.}\ \bibnamefont {Park}}, \bibinfo
  {author} {\bibfnamefont {A.}~\bibnamefont {Aspuru-Guzik}}, \bibinfo {author}
  {\bibfnamefont {D.~P.}\ \bibnamefont {Tabor}}, \bibinfo {author}
  {\bibfnamefont {J.}~\bibnamefont {Nelson}}, \bibinfo {author} {\bibfnamefont
  {J.}~\bibnamefont {Murphy}}, \bibinfo {author} {\bibfnamefont
  {A.}~\bibnamefont {Setlur}}, \bibinfo {author} {\bibfnamefont
  {J.}~\bibnamefont {Gregoire}}, \bibinfo {author} {\bibfnamefont
  {H.}~\bibnamefont {Li}}, \bibinfo {author} {\bibfnamefont {R.}~\bibnamefont
  {Xiao}}, \bibinfo {author} {\bibfnamefont {A.}~\bibnamefont {Ludwig}},
  \bibinfo {author} {\bibfnamefont {L.~W.}\ \bibnamefont {Martin}}, \bibinfo
  {author} {\bibfnamefont {A.~M.}\ \bibnamefont {Rappe}}, \bibinfo {author}
  {\bibfnamefont {S.-H.}\ \bibnamefont {Wei}}, \ and\ \bibinfo {author}
  {\bibfnamefont {J.}~\bibnamefont {Perkins}},\ }\href {\doibase
  10.1088/1361-6463/aad926} {\bibfield  {journal} {\bibinfo  {journal} {Journal
  of Physics D: Applied Physics}\ }\textbf {\bibinfo {volume} {52}},\ \bibinfo
  {pages} {013001} (\bibinfo {year} {2019})}\BibitemShut {NoStop}%
\bibitem [{\citenamefont {Luo}\ \emph {et~al.}(2021)\citenamefont {Luo},
  \citenamefont {Li}, \citenamefont {Wang}, \citenamefont {Faizan},\ and\
  \citenamefont {Zhang}}]{Luo2021}%
  \BibitemOpen
  \bibfield  {author} {\bibinfo {author} {\bibfnamefont {S.}~\bibnamefont
  {Luo}}, \bibinfo {author} {\bibfnamefont {T.}~\bibnamefont {Li}}, \bibinfo
  {author} {\bibfnamefont {X.}~\bibnamefont {Wang}}, \bibinfo {author}
  {\bibfnamefont {M.}~\bibnamefont {Faizan}}, \ and\ \bibinfo {author}
  {\bibfnamefont {L.}~\bibnamefont {Zhang}},\ }\href {\doibase
  10.1002/wcms.1489} {\bibfield  {journal} {\bibinfo  {journal} {WIREs
  Computational Molecular Science}\ }\textbf {\bibinfo {volume} {11}} (\bibinfo
  {year} {2021}),\ 10.1002/wcms.1489}\BibitemShut {NoStop}%
\bibitem [{\citenamefont {Curtarolo}\ \emph {et~al.}(2012)\citenamefont
  {Curtarolo}, \citenamefont {Setyawan}, \citenamefont {Hart}, \citenamefont
  {Jahnatek}, \citenamefont {Chepulskii}, \citenamefont {Taylor}, \citenamefont
  {Wang}, \citenamefont {Xue}, \citenamefont {Yang}, \citenamefont {Levy},
  \citenamefont {Mehl}, \citenamefont {Stokes}, \citenamefont {Demchenko},\
  and\ \citenamefont {Morgan}}]{Curtarolo2012}%
  \BibitemOpen
  \bibfield  {author} {\bibinfo {author} {\bibfnamefont {S.}~\bibnamefont
  {Curtarolo}}, \bibinfo {author} {\bibfnamefont {W.}~\bibnamefont {Setyawan}},
  \bibinfo {author} {\bibfnamefont {G.~L.}\ \bibnamefont {Hart}}, \bibinfo
  {author} {\bibfnamefont {M.}~\bibnamefont {Jahnatek}}, \bibinfo {author}
  {\bibfnamefont {R.~V.}\ \bibnamefont {Chepulskii}}, \bibinfo {author}
  {\bibfnamefont {R.~H.}\ \bibnamefont {Taylor}}, \bibinfo {author}
  {\bibfnamefont {S.}~\bibnamefont {Wang}}, \bibinfo {author} {\bibfnamefont
  {J.}~\bibnamefont {Xue}}, \bibinfo {author} {\bibfnamefont {K.}~\bibnamefont
  {Yang}}, \bibinfo {author} {\bibfnamefont {O.}~\bibnamefont {Levy}}, \bibinfo
  {author} {\bibfnamefont {M.~J.}\ \bibnamefont {Mehl}}, \bibinfo {author}
  {\bibfnamefont {H.~T.}\ \bibnamefont {Stokes}}, \bibinfo {author}
  {\bibfnamefont {D.~O.}\ \bibnamefont {Demchenko}}, \ and\ \bibinfo {author}
  {\bibfnamefont {D.}~\bibnamefont {Morgan}},\ }\href {\doibase
  10.1016/j.commatsci.2012.02.005} {\bibfield  {journal} {\bibinfo  {journal}
  {Computational Materials Science}\ }\textbf {\bibinfo {volume} {58}},\
  \bibinfo {pages} {218} (\bibinfo {year} {2012})}\BibitemShut {NoStop}%
\bibitem [{\citenamefont {Jain}\ \emph {et~al.}(2011)\citenamefont {Jain},
  \citenamefont {Hautier}, \citenamefont {Moore}, \citenamefont {Ping~Ong},
  \citenamefont {Fischer}, \citenamefont {Mueller}, \citenamefont {Persson},\
  and\ \citenamefont {Ceder}}]{Jain2011}%
  \BibitemOpen
  \bibfield  {author} {\bibinfo {author} {\bibfnamefont {A.}~\bibnamefont
  {Jain}}, \bibinfo {author} {\bibfnamefont {G.}~\bibnamefont {Hautier}},
  \bibinfo {author} {\bibfnamefont {C.~J.}\ \bibnamefont {Moore}}, \bibinfo
  {author} {\bibfnamefont {S.}~\bibnamefont {Ping~Ong}}, \bibinfo {author}
  {\bibfnamefont {C.~C.}\ \bibnamefont {Fischer}}, \bibinfo {author}
  {\bibfnamefont {T.}~\bibnamefont {Mueller}}, \bibinfo {author} {\bibfnamefont
  {K.~A.}\ \bibnamefont {Persson}}, \ and\ \bibinfo {author} {\bibfnamefont
  {G.}~\bibnamefont {Ceder}},\ }\href {\doibase
  10.1016/j.commatsci.2011.02.023} {\bibfield  {journal} {\bibinfo  {journal}
  {Computational Materials Science}\ }\textbf {\bibinfo {volume} {50}},\
  \bibinfo {pages} {2295} (\bibinfo {year} {2011})}\BibitemShut {NoStop}%
\bibitem [{\citenamefont {Huber}\ \emph {et~al.}(2020)\citenamefont {Huber},
  \citenamefont {Zoupanos}, \citenamefont {Uhrin}, \citenamefont {Talirz},
  \citenamefont {Kahle}, \citenamefont {H\"{a}uselmann}, \citenamefont
  {Gresch}, \citenamefont {M\"{u}ller}, \citenamefont {Yakutovich},
  \citenamefont {Andersen}, \citenamefont {Ramirez}, \citenamefont {Adorf},
  \citenamefont {Gargiulo}, \citenamefont {Kumbhar}, \citenamefont {Passaro},
  \citenamefont {Johnston}, \citenamefont {Merkys}, \citenamefont {Cepellotti},
  \citenamefont {Mounet}, \citenamefont {Marzari}, \citenamefont {Kozinsky},\
  and\ \citenamefont {Pizzi}}]{Huber2020}%
  \BibitemOpen
  \bibfield  {author} {\bibinfo {author} {\bibfnamefont {S.~P.}\ \bibnamefont
  {Huber}}, \bibinfo {author} {\bibfnamefont {S.}~\bibnamefont {Zoupanos}},
  \bibinfo {author} {\bibfnamefont {M.}~\bibnamefont {Uhrin}}, \bibinfo
  {author} {\bibfnamefont {L.}~\bibnamefont {Talirz}}, \bibinfo {author}
  {\bibfnamefont {L.}~\bibnamefont {Kahle}}, \bibinfo {author} {\bibfnamefont
  {R.}~\bibnamefont {H\"{a}uselmann}}, \bibinfo {author} {\bibfnamefont
  {D.}~\bibnamefont {Gresch}}, \bibinfo {author} {\bibfnamefont
  {T.}~\bibnamefont {M\"{u}ller}}, \bibinfo {author} {\bibfnamefont {A.~V.}\
  \bibnamefont {Yakutovich}}, \bibinfo {author} {\bibfnamefont {C.~W.}\
  \bibnamefont {Andersen}}, \bibinfo {author} {\bibfnamefont {F.~F.}\
  \bibnamefont {Ramirez}}, \bibinfo {author} {\bibfnamefont {C.~S.}\
  \bibnamefont {Adorf}}, \bibinfo {author} {\bibfnamefont {F.}~\bibnamefont
  {Gargiulo}}, \bibinfo {author} {\bibfnamefont {S.}~\bibnamefont {Kumbhar}},
  \bibinfo {author} {\bibfnamefont {E.}~\bibnamefont {Passaro}}, \bibinfo
  {author} {\bibfnamefont {C.}~\bibnamefont {Johnston}}, \bibinfo {author}
  {\bibfnamefont {A.}~\bibnamefont {Merkys}}, \bibinfo {author} {\bibfnamefont
  {A.}~\bibnamefont {Cepellotti}}, \bibinfo {author} {\bibfnamefont
  {N.}~\bibnamefont {Mounet}}, \bibinfo {author} {\bibfnamefont
  {N.}~\bibnamefont {Marzari}}, \bibinfo {author} {\bibfnamefont
  {B.}~\bibnamefont {Kozinsky}}, \ and\ \bibinfo {author} {\bibfnamefont
  {G.}~\bibnamefont {Pizzi}},\ }\href {\doibase 10.1038/s41597-020-00638-4}
  {\bibfield  {journal} {\bibinfo  {journal} {Scientific Data}\ }\textbf
  {\bibinfo {volume} {7}} (\bibinfo {year} {2020}),\
  10.1038/s41597-020-00638-4}\BibitemShut {NoStop}%
\bibitem [{\citenamefont {Feng}\ and\ \citenamefont
  {Cameron}(2007)}]{Feng2007}%
  \BibitemOpen
  \bibfield  {author} {\bibinfo {author} {\bibfnamefont {W.}~\bibnamefont
  {Feng}}\ and\ \bibinfo {author} {\bibfnamefont {K.}~\bibnamefont {Cameron}},\
  }\href {\doibase 10.1109/mc.2007.445} {\bibfield  {journal} {\bibinfo
  {journal} {Computer}\ }\textbf {\bibinfo {volume} {40}},\ \bibinfo {pages}
  {50} (\bibinfo {year} {2007})}\BibitemShut {NoStop}%
\bibitem [{\citenamefont {Feng}\ \emph {et~al.}(2008)\citenamefont {Feng},
  \citenamefont {Feng},\ and\ \citenamefont {Ge}}]{Feng2008}%
  \BibitemOpen
  \bibfield  {author} {\bibinfo {author} {\bibfnamefont {W.}~\bibnamefont
  {Feng}}, \bibinfo {author} {\bibfnamefont {X.}~\bibnamefont {Feng}}, \ and\
  \bibinfo {author} {\bibfnamefont {R.}~\bibnamefont {Ge}},\ }\href {\doibase
  10.1109/mitp.2008.8} {\bibfield  {journal} {\bibinfo  {journal} {IT
  Professional}\ }\textbf {\bibinfo {volume} {10}},\ \bibinfo {pages} {17}
  (\bibinfo {year} {2008})}\BibitemShut {NoStop}%
\bibitem [{\citenamefont {Herbst}\ and\ \citenamefont {Levitt}(2020)}]{ldos}%
  \BibitemOpen
  \bibfield  {author} {\bibinfo {author} {\bibfnamefont {M.~F.}\ \bibnamefont
  {Herbst}}\ and\ \bibinfo {author} {\bibfnamefont {A.}~\bibnamefont
  {Levitt}},\ }\href {\doibase 10.1088/1361-648x/abcbdb} {\bibfield  {journal}
  {\bibinfo  {journal} {Journal of Physics: Condensed Matter}\ } (\bibinfo
  {year} {2020}),\ 10.1088/1361-648x/abcbdb}\BibitemShut {NoStop}%
\bibitem [{\citenamefont {Canc\`{e}s}(2000)}]{Cances2000}%
  \BibitemOpen
  \bibfield  {author} {\bibinfo {author} {\bibfnamefont {E.}~\bibnamefont
  {Canc\`{e}s}},\ }\enquote {\bibinfo {title} {{SCF} algorithms for {HF}
  electronic calculations},}\ \ (\bibinfo  {publisher} {Springer Berlin
  Heidelberg},\ \bibinfo {year} {2000})\ pp.\ \bibinfo {pages}
  {17--43}\BibitemShut {NoStop}%
\bibitem [{\citenamefont {Canc\`{e}s}\ and\ \citenamefont
  {Le~Bris}(2000)}]{Cances2000a}%
  \BibitemOpen
  \bibfield  {author} {\bibinfo {author} {\bibfnamefont {E.}~\bibnamefont
  {Canc\`{e}s}}\ and\ \bibinfo {author} {\bibfnamefont {C.}~\bibnamefont
  {Le~Bris}},\ }\href {\doibase
  10.1002/1097-461X(2000)79:2<82::AID-QUA3>3.0.CO;2-I} {\bibfield  {journal}
  {\bibinfo  {journal} {International Journal of Quantum Chemistry}\ }\textbf
  {\bibinfo {volume} {79}},\ \bibinfo {pages} {82} (\bibinfo {year}
  {2000})}\BibitemShut {NoStop}%
\bibitem [{\citenamefont {Kudin}\ \emph {et~al.}(2002)\citenamefont {Kudin},
  \citenamefont {Scuseria},\ and\ \citenamefont {Canc\`{e}s}}]{Kudin2002}%
  \BibitemOpen
  \bibfield  {author} {\bibinfo {author} {\bibfnamefont {K.~N.}\ \bibnamefont
  {Kudin}}, \bibinfo {author} {\bibfnamefont {G.~E.}\ \bibnamefont {Scuseria}},
  \ and\ \bibinfo {author} {\bibfnamefont {E.}~\bibnamefont {Canc\`{e}s}},\
  }\href {\doibase 10.1063/1.1470195} {\bibfield  {journal} {\bibinfo
  {journal} {The Journal of Chemical Physics}\ }\textbf {\bibinfo {volume}
  {116}},\ \bibinfo {pages} {8255} (\bibinfo {year} {2002})},\ \Eprint
  {http://arxiv.org/abs/http://aip.scitation.org/doi/pdf/10.1063/1.1470195}
  {http://aip.scitation.org/doi/pdf/10.1063/1.1470195} \BibitemShut {NoStop}%
\bibitem [{\citenamefont {Francisco}\ \emph {et~al.}(2004)\citenamefont
  {Francisco}, \citenamefont {Mart{\i}nez},\ and\ \citenamefont
  {Mart{\i}nez}}]{francisco2004globally}%
  \BibitemOpen
  \bibfield  {author} {\bibinfo {author} {\bibfnamefont {J.~B.}\ \bibnamefont
  {Francisco}}, \bibinfo {author} {\bibfnamefont {J.~M.}\ \bibnamefont
  {Mart{\i}nez}}, \ and\ \bibinfo {author} {\bibfnamefont {L.}~\bibnamefont
  {Mart{\i}nez}},\ }\href@noop {} {\bibfield  {journal} {\bibinfo  {journal}
  {The Journal of chemical physics}\ }\textbf {\bibinfo {volume} {121}},\
  \bibinfo {pages} {10863} (\bibinfo {year} {2004})}\BibitemShut {NoStop}%
\bibitem [{\citenamefont {Francisco}\ \emph {et~al.}(2006)\citenamefont
  {Francisco}, \citenamefont {Mart{\'\i}nez},\ and\ \citenamefont
  {Mart{\'\i}nez}}]{francisco2006density}%
  \BibitemOpen
  \bibfield  {author} {\bibinfo {author} {\bibfnamefont {J.~B.}\ \bibnamefont
  {Francisco}}, \bibinfo {author} {\bibfnamefont {J.~M.}\ \bibnamefont
  {Mart{\'\i}nez}}, \ and\ \bibinfo {author} {\bibfnamefont {L.}~\bibnamefont
  {Mart{\'\i}nez}},\ }\href@noop {} {\bibfield  {journal} {\bibinfo  {journal}
  {Journal of Mathematical Chemistry}\ }\textbf {\bibinfo {volume} {40}},\
  \bibinfo {pages} {349} (\bibinfo {year} {2006})}\BibitemShut {NoStop}%
\bibitem [{\citenamefont {Canc\`{e}s}\ \emph {et~al.}(2021)\citenamefont
  {Canc\`{e}s}, \citenamefont {Kemlin},\ and\ \citenamefont
  {Levitt}}]{cances2020convergence}%
  \BibitemOpen
  \bibfield  {author} {\bibinfo {author} {\bibfnamefont {E.}~\bibnamefont
  {Canc\`{e}s}}, \bibinfo {author} {\bibfnamefont {G.}~\bibnamefont {Kemlin}},
  \ and\ \bibinfo {author} {\bibfnamefont {A.}~\bibnamefont {Levitt}},\ }\href
  {\doibase 10.1137/20m1332864} {\bibfield  {journal} {\bibinfo  {journal}
  {SIAM Journal on Matrix Analysis and Applications}\ }\textbf {\bibinfo
  {volume} {42}},\ \bibinfo {pages} {243} (\bibinfo {year} {2021})}\BibitemShut
  {NoStop}%
\bibitem [{\citenamefont {Marzari}\ \emph {et~al.}(1997)\citenamefont
  {Marzari}, \citenamefont {Vanderbilt},\ and\ \citenamefont
  {Payne}}]{marzari1997ensemble}%
  \BibitemOpen
  \bibfield  {author} {\bibinfo {author} {\bibfnamefont {N.}~\bibnamefont
  {Marzari}}, \bibinfo {author} {\bibfnamefont {D.}~\bibnamefont {Vanderbilt}},
  \ and\ \bibinfo {author} {\bibfnamefont {M.~C.}\ \bibnamefont {Payne}},\
  }\href@noop {} {\bibfield  {journal} {\bibinfo  {journal} {Physical Review
  Letters}\ }\textbf {\bibinfo {volume} {79}},\ \bibinfo {pages} {1337}
  (\bibinfo {year} {1997})}\BibitemShut {NoStop}%
\bibitem [{\citenamefont {Freysoldt}\ \emph {et~al.}(2009)\citenamefont
  {Freysoldt}, \citenamefont {Boeck},\ and\ \citenamefont
  {Neugebauer}}]{freysoldt2009direct}%
  \BibitemOpen
  \bibfield  {author} {\bibinfo {author} {\bibfnamefont {C.}~\bibnamefont
  {Freysoldt}}, \bibinfo {author} {\bibfnamefont {S.}~\bibnamefont {Boeck}}, \
  and\ \bibinfo {author} {\bibfnamefont {J.}~\bibnamefont {Neugebauer}},\
  }\href@noop {} {\bibfield  {journal} {\bibinfo  {journal} {Physical Review
  B}\ }\textbf {\bibinfo {volume} {79}},\ \bibinfo {pages} {241103} (\bibinfo
  {year} {2009})}\BibitemShut {NoStop}%
\bibitem [{\citenamefont {Marks}(2021)}]{Marks2021}%
  \BibitemOpen
  \bibfield  {author} {\bibinfo {author} {\bibfnamefont {L.~D.}\ \bibnamefont
  {Marks}},\ }\href {\doibase 10.1021/acs.jctc.1c00630} {\bibfield  {journal}
  {\bibinfo  {journal} {Journal of Chemical Theory and Computation}\ }\textbf
  {\bibinfo {volume} {17}},\ \bibinfo {pages} {5715} (\bibinfo {year}
  {2021})}\BibitemShut {NoStop}%
\bibitem [{\citenamefont {Marks}\ and\ \citenamefont
  {Luke}(2008)}]{marks2008robust}%
  \BibitemOpen
  \bibfield  {author} {\bibinfo {author} {\bibfnamefont {L.}~\bibnamefont
  {Marks}}\ and\ \bibinfo {author} {\bibfnamefont {D.}~\bibnamefont {Luke}},\
  }\href@noop {} {\bibfield  {journal} {\bibinfo  {journal} {Physical Review
  B}\ }\textbf {\bibinfo {volume} {78}},\ \bibinfo {pages} {075114} (\bibinfo
  {year} {2008})}\BibitemShut {NoStop}%
\bibitem [{\citenamefont {Gonze}(1996)}]{gonze1996towards}%
  \BibitemOpen
  \bibfield  {author} {\bibinfo {author} {\bibfnamefont {X.}~\bibnamefont
  {Gonze}},\ }\href@noop {} {\bibfield  {journal} {\bibinfo  {journal}
  {Physical Review B}\ }\textbf {\bibinfo {volume} {54}},\ \bibinfo {pages}
  {4383} (\bibinfo {year} {1996})}\BibitemShut {NoStop}%
\bibitem [{\citenamefont {Mermin}(1965)}]{mermin1965thermal}%
  \BibitemOpen
  \bibfield  {author} {\bibinfo {author} {\bibfnamefont {N.~D.}\ \bibnamefont
  {Mermin}},\ }\href@noop {} {\bibfield  {journal} {\bibinfo  {journal}
  {Physical Review}\ }\textbf {\bibinfo {volume} {137}},\ \bibinfo {pages}
  {A1441} (\bibinfo {year} {1965})}\BibitemShut {NoStop}%
\bibitem [{\citenamefont {Methfessel}\ and\ \citenamefont
  {Paxton}(1989)}]{methfessel1989high}%
  \BibitemOpen
  \bibfield  {author} {\bibinfo {author} {\bibfnamefont {M.}~\bibnamefont
  {Methfessel}}\ and\ \bibinfo {author} {\bibfnamefont {A.}~\bibnamefont
  {Paxton}},\ }\href@noop {} {\bibfield  {journal} {\bibinfo  {journal}
  {Physical Review B}\ }\textbf {\bibinfo {volume} {40}},\ \bibinfo {pages}
  {3616} (\bibinfo {year} {1989})}\BibitemShut {NoStop}%
\bibitem [{\citenamefont {Adler}(1962)}]{Adler1962}%
  \BibitemOpen
  \bibfield  {author} {\bibinfo {author} {\bibfnamefont {S.~L.}\ \bibnamefont
  {Adler}},\ }\href {\doibase 10.1103/physrev.126.413} {\bibfield  {journal}
  {\bibinfo  {journal} {Physical Review}\ }\textbf {\bibinfo {volume} {126}},\
  \bibinfo {pages} {413} (\bibinfo {year} {1962})}\BibitemShut {NoStop}%
\bibitem [{\citenamefont {Wiser}(1963)}]{Wiser1963}%
  \BibitemOpen
  \bibfield  {author} {\bibinfo {author} {\bibfnamefont {N.}~\bibnamefont
  {Wiser}},\ }\href {\doibase 10.1103/physrev.129.62} {\bibfield  {journal}
  {\bibinfo  {journal} {Physical Review}\ }\textbf {\bibinfo {volume} {129}},\
  \bibinfo {pages} {62} (\bibinfo {year} {1963})}\BibitemShut {NoStop}%
\bibitem [{\citenamefont {Dederichs}\ and\ \citenamefont
  {Zeller}(1983)}]{dederichs1983self}%
  \BibitemOpen
  \bibfield  {author} {\bibinfo {author} {\bibfnamefont {P.~H.}\ \bibnamefont
  {Dederichs}}\ and\ \bibinfo {author} {\bibfnamefont {R.}~\bibnamefont
  {Zeller}},\ }\href@noop {} {\bibfield  {journal} {\bibinfo  {journal}
  {Physical Review B}\ }\textbf {\bibinfo {volume} {28}},\ \bibinfo {pages}
  {5462} (\bibinfo {year} {1983})}\BibitemShut {NoStop}%
\bibitem [{\citenamefont {Kerker}(1981)}]{Kerker1981}%
  \BibitemOpen
  \bibfield  {author} {\bibinfo {author} {\bibfnamefont {G.~P.}\ \bibnamefont
  {Kerker}},\ }\href {\doibase 10.1103/physrevb.23.3082} {\bibfield  {journal}
  {\bibinfo  {journal} {Physical Review B}\ }\textbf {\bibinfo {volume} {23}},\
  \bibinfo {pages} {3082} (\bibinfo {year} {1981})}\BibitemShut {NoStop}%
\bibitem [{\citenamefont {Walker}\ and\ \citenamefont {Ni}(2011)}]{Walker2011}%
  \BibitemOpen
  \bibfield  {author} {\bibinfo {author} {\bibfnamefont {H.}~\bibnamefont
  {Walker}}\ and\ \bibinfo {author} {\bibfnamefont {P.}~\bibnamefont {Ni}},\
  }\href {\doibase 10.1137/10078356X} {\bibfield  {journal} {\bibinfo
  {journal} {SIAM Journal on Numerical Analysis}\ }\textbf {\bibinfo {volume}
  {49}},\ \bibinfo {pages} {1715} (\bibinfo {year} {2011})},\ \Eprint
  {http://arxiv.org/abs/https://doi.org/10.1137/10078356X}
  {https://doi.org/10.1137/10078356X} \BibitemShut {NoStop}%
\bibitem [{\citenamefont {Chupin}\ \emph {et~al.}(2020)\citenamefont {Chupin},
  \citenamefont {Dupuy}, \citenamefont {Legendre},\ and\ \citenamefont
  {S\'{e}r\'{e}}}]{Chupin2020}%
  \BibitemOpen
  \bibfield  {author} {\bibinfo {author} {\bibfnamefont {M.}~\bibnamefont
  {Chupin}}, \bibinfo {author} {\bibfnamefont {M.-S.}\ \bibnamefont {Dupuy}},
  \bibinfo {author} {\bibfnamefont {G.}~\bibnamefont {Legendre}}, \ and\
  \bibinfo {author} {\bibfnamefont {E.}~\bibnamefont {S\'{e}r\'{e}}},\ }\href
  {http://arxiv.org/abs/2002.12850v1} {\  (\bibinfo {year} {2020})},\ \Eprint
  {http://arxiv.org/abs/2002.12850v1} {2002.12850v1} \BibitemShut {NoStop}%
\bibitem [{\citenamefont {Saad}(2003)}]{Saad2003}%
  \BibitemOpen
  \bibfield  {author} {\bibinfo {author} {\bibfnamefont {Y.}~\bibnamefont
  {Saad}},\ }\href@noop {} {\emph {\bibinfo {title} {Iterative Methods for
  Sparse Linear Systems}}},\ \bibinfo {edition} {2nd}\ ed.,\ edited by\
  \bibinfo {editor} {\bibfnamefont {Y.}~\bibnamefont {Saad}}\ (\bibinfo
  {publisher} {SIAM Publishing},\ \bibinfo {year} {2003})\BibitemShut {NoStop}%
\bibitem [{\citenamefont {Kresse}\ and\ \citenamefont
  {Furthm\"{u}ller}(1996)}]{Kresse1996}%
  \BibitemOpen
  \bibfield  {author} {\bibinfo {author} {\bibfnamefont {G.}~\bibnamefont
  {Kresse}}\ and\ \bibinfo {author} {\bibfnamefont {J.}~\bibnamefont
  {Furthm\"{u}ller}},\ }\href {\doibase 10.1103/PhysRevB.54.11169} {\bibfield
  {journal} {\bibinfo  {journal} {Physical Review B}\ }\textbf {\bibinfo
  {volume} {54}},\ \bibinfo {pages} {11169} (\bibinfo {year}
  {1996})}\BibitemShut {NoStop}%
\bibitem [{\citenamefont {Herbst}\ \emph {et~al.}(2021)\citenamefont {Herbst},
  \citenamefont {Levitt},\ and\ \citenamefont {Cancès}}]{DFTKjcon}%
  \BibitemOpen
  \bibfield  {author} {\bibinfo {author} {\bibfnamefont {M.~F.}\ \bibnamefont
  {Herbst}}, \bibinfo {author} {\bibfnamefont {A.}~\bibnamefont {Levitt}}, \
  and\ \bibinfo {author} {\bibfnamefont {E.}~\bibnamefont {Cancès}},\ }\href
  {\doibase 10.21105/jcon.00069} {\bibfield  {journal} {\bibinfo  {journal}
  {Proc. JuliaCon Conf.}\ }\textbf {\bibinfo {volume} {3}},\ \bibinfo {pages}
  {69} (\bibinfo {year} {2021})}\BibitemShut {NoStop}%
\bibitem [{\citenamefont {Herbst}\ and\ \citenamefont
  {Levitt}(2021{\natexlab{a}})}]{DFTK}%
  \BibitemOpen
  \bibfield  {author} {\bibinfo {author} {\bibfnamefont {M.~F.}\ \bibnamefont
  {Herbst}}\ and\ \bibinfo {author} {\bibfnamefont {A.}~\bibnamefont
  {Levitt}},\ }\href {\doibase 10.5281/zenodo.5140420} {\enquote {\bibinfo
  {title} {{Density-functional toolkit (DFTK), version v0.3.10}},}\ } (\bibinfo
  {year} {2021}{\natexlab{a}}),\ \bibinfo {note}
  {\url{https://dftk.org}}\BibitemShut {NoStop}%
\bibitem [{\citenamefont {Perdew}\ \emph {et~al.}(1996)\citenamefont {Perdew},
  \citenamefont {Burke},\ and\ \citenamefont {Ernzerhof}}]{Perdew1996}%
  \BibitemOpen
  \bibfield  {author} {\bibinfo {author} {\bibfnamefont {J.~P.}\ \bibnamefont
  {Perdew}}, \bibinfo {author} {\bibfnamefont {K.}~\bibnamefont {Burke}}, \
  and\ \bibinfo {author} {\bibfnamefont {M.}~\bibnamefont {Ernzerhof}},\ }\href
  {\doibase 10.1103/PhysRevLett.77.3865} {\bibfield  {journal} {\bibinfo
  {journal} {Physical Review Letters}\ }\textbf {\bibinfo {volume} {77}},\
  \bibinfo {pages} {3865} (\bibinfo {year} {1996})}\BibitemShut {NoStop}%
\bibitem [{\citenamefont {Lehtola}\ \emph {et~al.}(2018)\citenamefont
  {Lehtola}, \citenamefont {Steigemann}, \citenamefont {Oliveira},\ and\
  \citenamefont {Marques}}]{Lehtola2018}%
  \BibitemOpen
  \bibfield  {author} {\bibinfo {author} {\bibfnamefont {S.}~\bibnamefont
  {Lehtola}}, \bibinfo {author} {\bibfnamefont {C.}~\bibnamefont {Steigemann}},
  \bibinfo {author} {\bibfnamefont {M.~J.}\ \bibnamefont {Oliveira}}, \ and\
  \bibinfo {author} {\bibfnamefont {M.~A.}\ \bibnamefont {Marques}},\ }\href
  {\doibase https://doi.org/10.1016/j.softx.2017.11.002} {\bibfield  {journal}
  {\bibinfo  {journal} {SoftwareX}\ }\textbf {\bibinfo {volume} {7}},\ \bibinfo
  {pages} {1 } (\bibinfo {year} {2018})}\BibitemShut {NoStop}%
\bibitem [{\citenamefont {Goedecker}\ \emph {et~al.}(1996)\citenamefont
  {Goedecker}, \citenamefont {Teter},\ and\ \citenamefont
  {Hutter}}]{Goedecker1996}%
  \BibitemOpen
  \bibfield  {author} {\bibinfo {author} {\bibfnamefont {S.}~\bibnamefont
  {Goedecker}}, \bibinfo {author} {\bibfnamefont {M.}~\bibnamefont {Teter}}, \
  and\ \bibinfo {author} {\bibfnamefont {J.}~\bibnamefont {Hutter}},\ }\href
  {\doibase 10.1103/PhysRevB.54.1703} {\bibfield  {journal} {\bibinfo
  {journal} {Physical Review B}\ }\textbf {\bibinfo {volume} {54}},\ \bibinfo
  {pages} {1703} (\bibinfo {year} {1996})}\BibitemShut {NoStop}%
\bibitem [{\citenamefont {Herbst}\ and\ \citenamefont
  {Levitt}(2021{\natexlab{b}})}]{reproducers}%
  \BibitemOpen
  \bibfield  {author} {\bibinfo {author} {\bibfnamefont {M.~F.}\ \bibnamefont
  {Herbst}}\ and\ \bibinfo {author} {\bibfnamefont {A.}~\bibnamefont
  {Levitt}},\ }\href@noop {} {\enquote {\bibinfo {title} {{Computational
  scripts and raw data for the presented numerical tests}},}\ } (\bibinfo
  {year} {2021}{\natexlab{b}}),\ \bibinfo {note}
  {\url{https://github.com/mfherbst/supporting-adaptive-damping}}\BibitemShut
  {NoStop}%
\bibitem [{\citenamefont {Bercx}\ and\ \citenamefont {Marzari}(2020)}]{tricky}%
  \BibitemOpen
  \bibfield  {author} {\bibinfo {author} {\bibfnamefont {M.}~\bibnamefont
  {Bercx}}\ and\ \bibinfo {author} {\bibfnamefont {N.}~\bibnamefont
  {Marzari}},\ }\href@noop {} {}\bibinfo {howpublished} {{Private
  communication}} (\bibinfo {year} {2020})\BibitemShut {NoStop}%
\bibitem [{\citenamefont {Winkelmann}\ \emph {et~al.}(2020)\citenamefont
  {Winkelmann}, \citenamefont {Di~Napoli}, \citenamefont {Wortmann},\ and\
  \citenamefont {Bl\"{u}gel}}]{Winkelmann2020}%
  \BibitemOpen
  \bibfield  {author} {\bibinfo {author} {\bibfnamefont {M.}~\bibnamefont
  {Winkelmann}}, \bibinfo {author} {\bibfnamefont {E.}~\bibnamefont
  {Di~Napoli}}, \bibinfo {author} {\bibfnamefont {D.}~\bibnamefont {Wortmann}},
  \ and\ \bibinfo {author} {\bibfnamefont {S.}~\bibnamefont {Bl\"{u}gel}},\
  }\href {\doibase 10.1103/physrevb.102.195138} {\bibfield  {journal} {\bibinfo
   {journal} {Physical Review B}\ }\textbf {\bibinfo {volume} {102}} (\bibinfo
  {year} {2020}),\ 10.1103/physrevb.102.195138}\BibitemShut {NoStop}%
\bibitem [{\citenamefont {Belkhouane}\ \emph {et~al.}(2015)\citenamefont
  {Belkhouane}, \citenamefont {Amari}, \citenamefont {Yakoubi}, \citenamefont
  {Tadjer}, \citenamefont {M\'{e}\c{c}abih}, \citenamefont {Murtaza},
  \citenamefont {Bin~Omran},\ and\ \citenamefont {Khenata}}]{Belkhouane2015}%
  \BibitemOpen
  \bibfield  {author} {\bibinfo {author} {\bibfnamefont {M.}~\bibnamefont
  {Belkhouane}}, \bibinfo {author} {\bibfnamefont {S.}~\bibnamefont {Amari}},
  \bibinfo {author} {\bibfnamefont {A.}~\bibnamefont {Yakoubi}}, \bibinfo
  {author} {\bibfnamefont {A.}~\bibnamefont {Tadjer}}, \bibinfo {author}
  {\bibfnamefont {S.}~\bibnamefont {M\'{e}\c{c}abih}}, \bibinfo {author}
  {\bibfnamefont {G.}~\bibnamefont {Murtaza}}, \bibinfo {author} {\bibfnamefont
  {S.}~\bibnamefont {Bin~Omran}}, \ and\ \bibinfo {author} {\bibfnamefont
  {R.}~\bibnamefont {Khenata}},\ }\href {\doibase 10.1016/j.jmmm.2014.10.094}
  {\bibfield  {journal} {\bibinfo  {journal} {Journal of Magnetism and Magnetic
  Materials}\ }\textbf {\bibinfo {volume} {377}},\ \bibinfo {pages} {211}
  (\bibinfo {year} {2015})}\BibitemShut {NoStop}%
\bibitem [{\citenamefont {Wollmann}\ \emph {et~al.}(2015)\citenamefont
  {Wollmann}, \citenamefont {Chadov}, \citenamefont {K\"{u}bler},\ and\
  \citenamefont {Felser}}]{Wollmann2015}%
  \BibitemOpen
  \bibfield  {author} {\bibinfo {author} {\bibfnamefont {L.}~\bibnamefont
  {Wollmann}}, \bibinfo {author} {\bibfnamefont {S.}~\bibnamefont {Chadov}},
  \bibinfo {author} {\bibfnamefont {J.}~\bibnamefont {K\"{u}bler}}, \ and\
  \bibinfo {author} {\bibfnamefont {C.}~\bibnamefont {Felser}},\ }\href
  {\doibase 10.1103/physrevb.92.064417} {\bibfield  {journal} {\bibinfo
  {journal} {Physical Review B}\ }\textbf {\bibinfo {volume} {92}} (\bibinfo
  {year} {2015}),\ 10.1103/physrevb.92.064417}\BibitemShut {NoStop}%
\bibitem [{\citenamefont {Shi}\ \emph {et~al.}(2020)\citenamefont {Shi},
  \citenamefont {Li}, \citenamefont {Zhang}, \citenamefont {Zhai},
  \citenamefont {Jiang}, \citenamefont {Wang}, \citenamefont {Chen},
  \citenamefont {Yan}, \citenamefont {Zhang},\ and\ \citenamefont
  {Liu}}]{Shi2020}%
  \BibitemOpen
  \bibfield  {author} {\bibinfo {author} {\bibfnamefont {B.}~\bibnamefont
  {Shi}}, \bibinfo {author} {\bibfnamefont {J.}~\bibnamefont {Li}}, \bibinfo
  {author} {\bibfnamefont {C.}~\bibnamefont {Zhang}}, \bibinfo {author}
  {\bibfnamefont {W.}~\bibnamefont {Zhai}}, \bibinfo {author} {\bibfnamefont
  {S.}~\bibnamefont {Jiang}}, \bibinfo {author} {\bibfnamefont
  {W.}~\bibnamefont {Wang}}, \bibinfo {author} {\bibfnamefont {D.}~\bibnamefont
  {Chen}}, \bibinfo {author} {\bibfnamefont {Y.}~\bibnamefont {Yan}}, \bibinfo
  {author} {\bibfnamefont {G.}~\bibnamefont {Zhang}}, \ and\ \bibinfo {author}
  {\bibfnamefont {P.-F.}\ \bibnamefont {Liu}},\ }\href {\doibase
  10.1039/d0cp03226c} {\bibfield  {journal} {\bibinfo  {journal} {Physical
  Chemistry Chemical Physics}\ }\textbf {\bibinfo {volume} {22}},\ \bibinfo
  {pages} {23185} (\bibinfo {year} {2020})}\BibitemShut {NoStop}%
\bibitem [{\citenamefont {He}\ \emph {et~al.}(2018)\citenamefont {He},
  \citenamefont {Naghavi}, \citenamefont {Hegde}, \citenamefont {Amsler},\ and\
  \citenamefont {Wolverton}}]{He2018}%
  \BibitemOpen
  \bibfield  {author} {\bibinfo {author} {\bibfnamefont {J.}~\bibnamefont
  {He}}, \bibinfo {author} {\bibfnamefont {S.~S.}\ \bibnamefont {Naghavi}},
  \bibinfo {author} {\bibfnamefont {V.~I.}\ \bibnamefont {Hegde}}, \bibinfo
  {author} {\bibfnamefont {M.}~\bibnamefont {Amsler}}, \ and\ \bibinfo {author}
  {\bibfnamefont {C.}~\bibnamefont {Wolverton}},\ }\href {\doibase
  10.1021/acs.chemmater.8b01096} {\bibfield  {journal} {\bibinfo  {journal}
  {Chemistry of Materials}\ }\textbf {\bibinfo {volume} {30}},\ \bibinfo
  {pages} {4978} (\bibinfo {year} {2018})}\BibitemShut {NoStop}%
\bibitem [{\citenamefont {Jiang}\ and\ \citenamefont {Yang}(2021)}]{Jiang2021}%
  \BibitemOpen
  \bibfield  {author} {\bibinfo {author} {\bibfnamefont {S.}~\bibnamefont
  {Jiang}}\ and\ \bibinfo {author} {\bibfnamefont {K.}~\bibnamefont {Yang}},\
  }\href {\doibase 10.1016/j.jallcom.2021.158854} {\bibfield  {journal}
  {\bibinfo  {journal} {Journal of Alloys and Compounds}\ }\textbf {\bibinfo
  {volume} {867}},\ \bibinfo {pages} {158854} (\bibinfo {year}
  {2021})}\BibitemShut {NoStop}%
\bibitem [{\citenamefont {Hager}\ and\ \citenamefont
  {Zhang}(2006)}]{hager2006survey}%
  \BibitemOpen
  \bibfield  {author} {\bibinfo {author} {\bibfnamefont {W.~W.}\ \bibnamefont
  {Hager}}\ and\ \bibinfo {author} {\bibfnamefont {H.}~\bibnamefont {Zhang}},\
  }\href@noop {} {\bibfield  {journal} {\bibinfo  {journal} {Pacific Journal of
  Optimization}\ }\textbf {\bibinfo {volume} {2}},\ \bibinfo {pages} {35}
  (\bibinfo {year} {2006})}\BibitemShut {NoStop}%
\bibitem [{\citenamefont {Higham}(2008)}]{higham2008functions}%
  \BibitemOpen
  \bibfield  {author} {\bibinfo {author} {\bibfnamefont {N.~J.}\ \bibnamefont
  {Higham}},\ }\href@noop {} {\emph {\bibinfo {title} {Functions of matrices:
  theory and computation}}}\ (\bibinfo  {publisher} {SIAM},\ \bibinfo {year}
  {2008})\BibitemShut {NoStop}%
\bibitem [{\citenamefont {Levitt}(2020)}]{levitt2020screening}%
  \BibitemOpen
  \bibfield  {author} {\bibinfo {author} {\bibfnamefont {A.}~\bibnamefont
  {Levitt}},\ }\href@noop {} {\bibfield  {journal} {\bibinfo  {journal}
  {Archive for Rational Mechanics and Analysis}\ }\textbf {\bibinfo {volume}
  {238}},\ \bibinfo {pages} {901} (\bibinfo {year} {2020})}\BibitemShut
  {NoStop}%
\bibitem [{\citenamefont {Levitt}(2012)}]{levitt2012convergence}%
  \BibitemOpen
  \bibfield  {author} {\bibinfo {author} {\bibfnamefont {A.}~\bibnamefont
  {Levitt}},\ }\href@noop {} {\bibfield  {journal} {\bibinfo  {journal} {ESAIM:
  Mathematical Modelling and Numerical Analysis}\ }\textbf {\bibinfo {volume}
  {46}},\ \bibinfo {pages} {1321} (\bibinfo {year} {2012})}\BibitemShut
  {NoStop}%
\bibitem [{\citenamefont {Lojasiewicz}(1965)}]{lojasiewicz1965ensembles}%
  \BibitemOpen
  \bibfield  {author} {\bibinfo {author} {\bibfnamefont {S.}~\bibnamefont
  {Lojasiewicz}},\ }\href@noop {} {\bibfield  {journal} {\bibinfo  {journal}
  {Lectures Notes IHES (Bures-sur-Yvette)}\ } (\bibinfo {year}
  {1965})}\BibitemShut {NoStop}%
\end{thebibliography}%
\end{document}